\documentclass[10pt,a4paper]{article}             % e-submission use XeLatex

\usepackage{amsmath,amsthm,amssymb,mathrsfs,bm,bbm,enumerate,physics}
\usepackage{tikz}
\usetikzlibrary{intersections,calc,arrows.meta}
\usepackage{float,placeins,adjustbox}
\usepackage[hyphens]{url}
\usepackage{hyperref}
\usepackage{cleveref}

\usepackage{comment}
\usepackage{caption}
\newcommand\myshade{75}
\colorlet{mylinkcolor}{red}
\colorlet{mycitecolor}{green}
\colorlet{myurlcolor}{blue}
\hypersetup{colorlinks=true, anchorcolor=cyan, 
linkcolor=mylinkcolor!\myshade!black, 
citecolor=mycitecolor!\myshade!black,
urlcolor=myurlcolor!\myshade!black } 

\newcommand{\beq}{\begin{equation}}
\newcommand{\eeq}{\end{equation}}
\newcommand{\nn}{\nonumber}
\newcommand{\ra}{\rightarrow}
\newcommand{\lra}{\leftrightarrow}
\newcommand{\qqi}{q-q^{-1}}
\newcommand{\eps}{\epsilon}
\newcommand{\scz}{\widehat{\mkern-4mu CZ}}

\newcommand{\Op}[3]{ \hat{#1}_{#2}^{({#3})}}

\newcommand{\Exp}[1]{\exp\left({#1}\right)}
\newcommand{\Tm}[1]{{\mathscr T}_{\bm{#1} } }

\newcommand\delt{\partial_{\theta}}
\newcommand\delx{\partial_x}          

\theoremstyle{definition}
\newtheorem*{rmk*}{$\ddagger$  Remark}

\theoremstyle{remark}
\newtheorem*{exm*}{$\spadesuit$ Example}

\makeatletter
\long\def\@makefntext#1{\parindent 1em\noindent
\@hangfrom{\hbox to 1.8em{\hss$^{\@thefnmark}$}}#1}
\makeatother
\newif\ifcolor
\colorfalse  %  <--  choose true or false
\ifcolor
  \def\red#1{{\color{red}#1}}
  
\else
  \def\red#1{#1}
  
\fi

\begin{document}
\topmargin 0pt
\oddsidemargin 0mm
\renewcommand{\thefootnote}{\fnsymbol{footnote}}

%\begin{flushright}
%manuscript 0.5 \\
%CZ-alg-003
%\end{flushright}
\vspace*{0.5cm}

\begin{center}
{\Large Quantum Superspace and Bloch Electron Systems with Zeeman Effects: \\
$\ast$-Bracket Formalism for Super Curtright-Zachos Algebras}
\vspace{1.5cm}

{\large Haru-Tada Sato${\,}^{a,b}$
\footnote{\,\,Corresponding author.  E-mail address: satoh@isfactory.co.jp \\
${}^{a}$\,\,This research was completed during the author's tenure as a visiting researcher there, ending on October 31, 2024.}
}\\

{${}^{a}$\em Department of Physics, Graduate School of Science, \\Osaka Metropolitan University \\
Nakamozu Campus, Sakai, Osaka 599-8531, Japan}\\

{${}^{b}$\em Department of Data Science, i's Factory Corporation, Ltd.\\
     Kanda-nishiki-cho 2-7-6, Tokyo 101-0054, Japan}\\
%
%\vspace{0.5cm}
%This article is registered under preprint number: /hep-th/*******'.
\end{center}

\vspace{0.1cm}

\abstract{
We introduce supersymmetric extensions of the Hom-Lie deformation of the Virasoro algebra (super Curtright-Zachos algebra), as realized in the GL(1,1) quantum superspace, for Bloch electron systems under Zeeman effects. By examining the duality inherent in quantum superspace scaling operators, we establish a correspondence between quantum superspace and its physical realization through a novel operator mixing mechanism. For the continuous case, we construct super Curtright-Zachos algebra using magnetic translations and spin matrix bases, demonstrating explicit realizations for both $N=1$ and $N=2$ supersymmetric algebras with a natural $N=2$ decomposition. For the discrete case, we establish cyclic matrix representations in tight-binding models. We organize these structures through the $\ast$-bracket formalism with $Z_2$-grading, revealing how the quantum superspace structure manifests in physical systems while preserving essential algebraic properties.}
\vspace{0.5cm}

\begin{description}
\item[Keywords:] quantum superspace, Hom-Lie deformation, Virasoro algebra, magnetic translation, Zeeman effect, tight binding model 
\item[MSC:] 17B61, 17B68, 81R50, 81R60
%\item[arXiv]: ****.****
\end{description}

%\footnote{}
%
%
%%%%%%%%%%%%%%%%%%%%  INDEX  %%%%%%%%%%%%%%%%%%%%%%%%%%%%%%%%%%%%%%%%
%\newpage
%\tableofcontents
%\setcounter{page}{1}
%\renewcommand{\thepage}{\roman{page}}

%%%%%%%%%%%%%%%%%%%%%%%%%%%%%%%%%%%%%%%%%%%%%%%%%%%%%%%%%%%%%%%%%%%%%
%                           1.     Introduction
%
%%%%%%%%%%%%%%%%%%%%%%%%%%%%%%%%%%%%%%%%%%%%%%%%%%%%%%%%%%%%%%%%%%%%%
\newpage
\setcounter{page}{1}
\setcounter{footnote}{0}
\setcounter{equation}{0}
\setcounter{secnumdepth}{4}
\renewcommand{\thepage}{\arabic{page}}
\renewcommand{\thefootnote}{\arabic{footnote})}
\renewcommand{\theequation}{\thesection.\arabic{equation}}

\section{Introduction}
\subsection{background}
\label{sec:BG}
\indent

Noncommutative geometry is an interesting subject in the field of physics. 
It is considered to be induced by strong external fields such as magnetic fields, 
and it covers a wide range of topics including the theory of noncommutative fields~\cite{NCFT,SW}, quantum Hall states~\cite{latestQH}, 
AdS/CFT and black holes~\cite{AdS1}-\cite{Edge3}. Moyal deformations of self-dual gravity 
have been studied in the context of noncommutativity and infinite-dimensional symmetries~\cite{BHS}-\cite{Strom3}.

There are two main tools for describing noncommutative space physics: noncommutative spaces by Moyal deformation~\cite{Moyal1,Moyal2} and noncommutative spaces conjugate to quantum groups~\cite{ma1}-\cite{maj2}. These theories have clear differences in their background and the nature of noncommutativity they handle. Fundamentally, Moyal deformation is primarily a method to introduce noncommutativity of coordinates in flat space or phase space, assuming classical symmetry and group structures. Through the Moyal product, it describes field theories and quantum mechanical actions on noncommutative spaces.

To illustrate the structure of Moyal deformation more concretely, let us consider the Moyal sine algebra, which is also known as the FFZ algebra~\cite{FFZ1}-\cite{FFZ3}. This algebra emerges from the Moyal bracket deformation, which provides a Lie-algebraic deformation of the Poisson brackets. The Moyal bracket and its star product are defined as follows~\cite{Moyal1,Moyal2}: 
\begin{align}
& \{f(x,p),g(x,p)\}_\ast =\frac{2}{\hbar}\sin(\frac{\hbar}{2}\theta^{ab}\partial^a_1\partial^b_2)f(x,p)g(x,p)\,,\\
&f*g=\exp{i\frac{\hbar}{2}\theta^{ab} \partial_1^a \partial_2^b }f(x,p)g(x,p)\,,
\label{f*g}
\end{align}
where $\theta^{xp}=-\theta^{px}=-\omega$, and $\partial_1$ and $\partial_2$ denote 
forward (left) and backward (right) derivative operations, respectively. 
The Moyal quantization leads to the $SU(\infty)$ Lie algebra, so-called 
the Moyal sine algebra
\beq
[\tau_{n,k}, \tau_{m,l}] = 2i \sine(\frac{\hbar\omega}{2}(nl-mk)) \tau_{n+m,k+l} \,, \label{Msine}
\eeq
if one takes the basis of $T^2$ phase space
\beq
\tau_{n,k}=e^{i(nx+kp)} \,. \label{txp}
\eeq
Typical realization of the algebra is the magnetic translations (MT)~\cite{Zak1,Zak2}, and 
the hyperbolic sine version of the algebra also appears in the context of quantum Hall physics~\cite{GMP}.

On the other hand, noncommutative spaces conjugate to quantum groups reflect spaces with quantum group symmetry, where the coordinates and actions are quantized according to quantum groups. Quantum groups~\cite{ma1}-\cite{maj2} are quantizations of Lie groups and Lie algebras~\cite{DJ}-\cite{wor3} that describe conventional symmetries. Specifically, they introduce symmetries handled within the framework of noncommutative algebras by generalizing classical symmetries based on Lie groups and algebras.

In quantum group conjugate noncommutative spaces, not only do coordinates exhibit noncommutativity, but they also manifest noncommutative algebraic structures reflecting the nontrivial Hopf algebra structure of quantum groups. This results in a more sophisticated and complex structured noncommutativity than merely coordinate noncommutativity, as both spatial symmetries and group actions are quantized simultaneously. In this sense, the space symmetry of Moyal deformation can be considered classical.

These two approaches to noncommutativity - Moyal deformation and quantum groups - have traditionally been studied separately. The simplest quantum group covariant space (or simply, quantum space (QS)) is the two-dimensional quantum plane satisfying the relation $xy=qyx$, where differential operations involve $q$-derivative operators such as $\partial_x x=1+q^{n}x\partial_x$. The value of $n$ varies depending on the quantum group, for example, $n=2$ for $GL_q(2)$~\cite{wz}, $-2$ for $GL_q(1,1)$~\cite{glq1,glq2,glq3}, $-1$ for quantum affine transformation~\cite{affine1,affine2}. 

Interestingly, there exists an algebra that exhibits characteristics of both approaches: the Curtright-Zachos (CZ) algebra, which can be constructed using the $q$-derivative operators of quantum spaces while showing properties reminiscent of Moyal deformation~\cite{AS3}. This algebra has emerged as a fascinating bridge connecting the noncommutativity of Moyal deformation and quantum planes~\cite{AS2,AS2e}, suggesting deeper connections between these seemingly distinct approaches to noncommutative geometry.

%%%%%%%%%%%%%%%%%%%%%%%%%%%%%%
%   1.2    CZ algebras
%%%%%%%%%%%%%%%%%%%%%%%%%%%%%%
\subsection{CZ algebras}

The CZ algebra was proposed by Curtright and Zachos as a 
$q$-deformation of the Virasoro algebra~\cite{CZ}, 
\beq
[L_n,L_m]_\ast=(L_nL_m)_\ast-(L_mL_n)_\ast=[n-m]L_{n+m}\,,  \label{CZ} 
\eeq
where $(L_nL_m)_\ast=q^{m-n}L_nL_m$ and the $q$-bracket symbol $[A]$ is defined as
\beq
[A]=\frac{q^A -q^{-A}}{\qqi}\,,\quad  \mbox{where}\quad q=e^{i\hbar\omega} \,, 
\label{qbraA}
\eeq
This is mathematically interpreted as a Hom-Lie algebra~\cite{Hom1,Hom2,Hom3}, 
which is why we refer to the CZ algebra as the Hom-Lie-Virasoro algebra. 
In addition to the original algebra \eqref{CZ} (which we denote as $CZ^+$), there are two other variations: $CZ^-$ which is obtained by $q$-inversion ($q\ra q^{-1}$), and $CZ^\ast$ which unifies these algebras~\cite{AS2,AS2e}.

There are many interesting results concerning the CZ algebras, including 
central extensions and operator product formula (OPE)~\cite{AS,Poisson}, 
$q$-harmonic oscillators~\cite{CKL,CKL2}, matrix representations~\cite{NQ,HHTZ2}, 
and fractional spin representations~\cite{MZ}. 
Supersymmetric extensions~\cite{MZ,superCZ,superCZ2,super3,HHT} and 
multi-parameter deformations~\cite{pqCZ}-\cite{pqCZ2} are also studied. 
Recently, deformaton of open string filed theory has been investigated~\cite{qstr}. 
(see also Section 1.3 in \cite{AS2} for further references.)

The generators of $CZ$ can be expressed in terms of MT operators satisfying FFZ algebra or their corresponding cyclic matrices~\cite{NQ}, and the MT operators behave like a quantum plane as noncommutative translation operators. Through this connection between magnetic translations and quantum planes, several fundamental aspects of noncommutative geometry have recently become clear~\cite{AS3,AS2}.

The phase factors in the $\ast$-bracket $(L_nL_m)_\ast$ can be understood as 
the phase differences generated according to the paths traced by MT operations~\cite{AS2,AS2e}. The operational behavior of MT's matrix representation reveals that the phase shift arising from the path dependence of translational operations between two points on the plane leads to the quantum space noncommutativity of the TBM in a square lattice space~\cite{AS2,AS2e}. The phase-shifted commutation relations \eqref{CZ} can be explained by this path-dependent phase difference based on Weyl matrix representations. In other words, introducing phase-shifting products and changing the commutators to phase-shifted ones as quantum plane effects lead to the derivation of the Hom-Lie-Virasoro algebra. 

When MT is expressed as angular momentum representation on a cylinder, $L_n$ is represented as a one-dimensional $q$-differential operator, and this $q$-differential operator satisfies the same commutation relations as the differential operators on quantum planes that behave covariantly with quantum groups. While the relationship with truly quantum group covariant multidimensional spaces remains unclear, the one-dimensional reduction clearly shows the same structure. We can therefore appropriately call it a quantum line, which represents the simplest form of quantum space.

Furthermore, since MT follows FFZ algebra, the $CZ$ algebra is expected to have a Moyal structure. Indeed, it has been shown that the $CZ$ algebra $\mathcal{CZ}$, when extended by scaling operators, possesses a Moyal $\ast$-product structure~\cite{AS3}. 
$\mathcal{CZ}$ includes $CZ$ as a subalgebra, and the scaling operator structure between $\mathcal{CZ}$ and $CZ$ takes exactly the same form as the internal scaling operator structure of MT operators. However, a mystery remains: CZ requires doubling the exchange phase of $q$ compared to $\mathcal{CZ}$ to take the Moyal structure. While we have not yet fully revealed the complete picture of $CZ$, we seem to be approaching the solution.

This example suggests that by considering extensions, we can grasp unknown (or desired) properties of the original system. The significance lies in gaining guidance on where to embed these properties by utilizing the broader symmetries and degrees of freedom possessed by the extended object.

While attempts have been made to construct CZ algebra using quantum planes covariant under quantum groups~\cite{superCZ,superCZ2,HHT,HHTZ} (particularly in the context of supersymmetry), a concrete physical realization in Bloch electron systems under Zeeman effects has been recently achieved~\cite{SCZ4}. This provides a foundation for our current investigation into a novel theoretical structure where bosonic and fermionic operators exhibit intricate mixing behaviors in the quantum superspace correspondence. For this purpose, a promising research direction is to investigate the supersymmetric extension of physical systems~\cite{AS3,AS2,AS2e} that exhibit CZ algebra.

Fortunately, in the quantum group approach, quantum superspaces (QSS) covariant under $Osp_q(1,2)$ and $GL_q(1,1)$ have been studied, and several super CZ algebras have been obtained~\cite{superCZ,superCZ2}. Therefore, it is convenient to study the super CZ algebra to investigate how our super CZ is related to two-dimensional electron systems. There exists another type of super CZ~\cite{MZ,super3,HHT} that has anticommutative supercharge with quite simple quantum superspace relations, specified as 2-parameter~\cite{HHT}. However, since this is not suitable for our purpose, we will not discuss it further in this paper.

%%%%%%%%%%%%%%%%%%%%%%%%%%%%%%
%   1.3    Contents
%%%%%%%%%%%%%%%%%%%%%%%%%%%%%%
\subsection{Focus}
%\label{sec:}
\indent

In this paper, we investigate how CZ algebra supersymmetrization manifests in physical systems exhibiting supersymmetry. CZ algebra emerges naturally in two-dimensional electron systems under strong magnetic fields. Through magnetic translation operator realizations and cyclic matrix representations, this algebra has been shown to be closely connected to the quantum plane structure describing noncommutative geometry~\cite{AS2}.
While the basic construction of supersymmetric CZ algebra through MSB (magnetic translations and spin matrix bases) for $N=1$ and $N=2$ has been established~\cite{SCZ4}, our focus is on presenting the detailed calculations and mathematical structures that were omitted in that work, thereby providing a complete theoretical foundation for understanding supersymmetric CZ algebras in physical systems.

Super CZ algebras have been constructed using quantum superspace (QSS) noncommutativity, with at least three types of algebras known~\cite{superCZ,superCZ2}. Recent work has demonstrated their physical realization in Bloch electron systems under Zeeman effects~\cite{SCZ4}. Building upon these results, we present a more comprehensive understanding through the analysis of bosonic and fermionic operator correspondence in the MSB-QSS framework and their role in the supersymmetric structures.

The noncommutative structures of MTs and quantum space are remarkably similar, and in special cases of quantum space, they are known to coincide. With quantum space analogy providing new guidance, the possibility of considering superalgebra as more tangible has expanded. Therefore, the main focus of this paper is how super CZ algebra constructed by QSS connects with the supersymmetrization of CZ generators constructed by MT through the Zeeman effect.

This paper is organized as follows. In Section~\ref{sec:CZ}, we provide a brief overview of the CZ algebra and MT operators from the previous paper, and explain the necessary notation. In Section~\ref{sec:QSS}, we summarize three types of super CZ algebras 
based on QSS formalism. Starting with the continuous case in Section~\ref{sec:MSB}, 
we develop our theoretical framework of mixing mechanism, which is necessary for the MSB-QSS correspondence. 
In Section~\ref{sec:*form}, we establish a possible formulation of super CZ algebras in the unified $\ast$-bracket framework with Z$_2$-grading structure. We encounter 
a different $\ast$-bracket structure between $N=1$ and 2. 
In Section~\ref{sec:TBM}, we extend this framework to the discrete case and present explicit matrix representations.
Finally, in Section~\ref{sec:end}, we summarize our findings and discuss several open questions and future perspectives. 

\red{
For the reader's convenience, Appendix~\ref{sec:def} collects the notation and conventions used throughout the paper. 
Appendix~\ref{sec:3types} provides a systemtic review of the three types of super CZ algebras. 
Definitions of operators are summarized in 
Appendix~\ref{app:MSB} for the MT and MSB representations, and in Appendix~\ref{app:Weyl} for the TBM discrete system (Weyl matrix representation). 
Appendix~\ref{sec:SV} summarizes the SSM (superspace and spin matrix) correspondence and the super Virasoro algebra in the electron spin system under a static magnetic field.
}
%%%%%%%%%%%%%%%%%%%%%%%%%%%%%%%%%%%%%%%%%%%%%%%%%%%%%%%%%%%%%%%%%%
%            Section 2.    CZ algebra
%
%%%%%%%%%%%%%%%%%%%%%%%%%%%%%%%%%%%%%%%%%%%%%%%%%%%%%%%%%%%%%%%%%%
\setcounter{equation}{0}
\section{Magnetic Translation (MT) and $CZ$ Algebras}\label{sec:CZ}
\indent

The CZ algebra \eqref{CZ} is a Hom-Lie deformation of the Virasoro algebra, and it 
is shown to be obtained by the FFZ generators~\cite{AS2,AS2e}. This algebra, denoted as 
$CZ^+$, has two related algebras: $CZ^-$ and $CZ^\ast$. The algebras $CZ^\pm$ are symmetric in the interchange of $q\lra q^{-1}$, and the $CZ^\ast$ algebra is an extended algebra composed of $CZ^\pm$.

What we call the FFZ algebra here was originally given by \eqref{Msine} with the 
introduction of the deformation parameter $q$ and the $q$-bracket defined in \eqref{qbraA}. Changing the normalization
\beq
T_n^{(k)} =\frac{1}{\qqi} \tau_{n,k}  \label{T2base}
\eeq
we have the FFZ algebra in the $q$-bracket form
\beq
[T_n^{(k)}\,, T_m^{(l)}]=[\frac{nl-mk}{2}]T_{n+m}^{(k+l)}  \,,  \label{eq:FFZ}
\eeq
where $\tau_{n,k}$ can be generalized to any operator $t_n^{(k)}$ that satisfies the Moyal star product relations~\cite{FFZ1}-\cite{FFZ3}:
\beq
t_n^{(k)}t_m^{(l)}= q^{\frac{nl-mk}{2}}  t_{n+m}^{(k+l)}\,. \label{trans}
\eeq
From this, we obtain the exchange relation 
\beq
  t_n^{(k)}t_m^{(l)}= q^{nl-mk}  t_m^{(l)}t_n^{(k)} \,.
\eeq
$T_n^{(k)}$ also satisfies the same exchange relation by definition 
$t_n^{(k)}:=(q-q^{-1})T_n^{(k)}$. These operators satisfy the fusion rule, which provides a realization of the FFZ algebra \eqref{eq:FFZ}:
\beq
T_n^{(k)}T_m^{(l)}= \frac{1}{\qqi} q^{\frac{nl-mk}{2}}  T_{n+m}^{(k+l)}\,.  \label{Trans}
\eeq
Magnetic translation (MT) operators have been demonstrated to satisfy these 
fusion and exchange relations~\cite{Zak1,Zak2}. Here, the normalization factor $\qqi$ 
facilitates the connection between the FFZ algebra \eqref{eq:FFZ} and CZ algebra in the 
regime $q\not=1$.

In this paper, we deal with the angular momentum representation instead of \eqref{txp}, where the MT operator $\tau_{n,k}$ is given by $\Op{t}{n}{k}$ as follows: 
\begin{eqnarray}
\Op{t}{n}{k} &=&
z^n q^{-k(z\partial +\frac{n}{2}+\Delta)} =e^{n\ln{z}}e^{-ia^2l_B^{-2}k(z\partial +\frac{n}{2}+\Delta)} \nn\\
&=&\Exp{\frac{i}{\hbar}\bm{\lambda}\cdot\bm{\Phi}}\,,  \label{Tmtheta}
\end{eqnarray}
where the vectors $\bm{\lambda}$ and $\bm{\Phi}$ are defined by
\beq
\bm{\lambda}=(nl_B,k\frac{a^2}{l_B})\,,
\eeq
\beq
(\Phi_1,\Phi_2)=(\frac{\hbar}{l_B}\varphi, -\frac{1}{l_B}\mathcal{J}_3 
-\frac{\hbar}{l_B}\Delta)\,,\quad\, \mathcal{J}_3=-i\hbar\partial_\varphi\,,\quad
z=e^{i\varphi}\,. \label{varset}
\eeq
Here, $l_B=\sqrt{\hbar c/eB}$ represents the magnetic length characteristic of the system, and $a$ denotes a unit length scale. Through these parameters, the deformation parameter $q$ is naturally defined as
\beq
q=\exp{ia^2l_B^{-2}}=e^{i\hbar\omega}\,.
\eeq 
The phase space of $\Op{t}{n}{k}$ is characterized by the operators $\mathcal{J}_3$ and $\varphi$, which satisfy the fundamental commutation relations
\beq
[\mathcal{J}_3,\varphi]=-i\hbar\,,\quad [\Phi_i,\Phi_j]=-i\frac{\hbar^2}{l_B^2}\eps^{ij}\,.
\label{TTcom}
\eeq
In this framework, the Moyal product \eqref{f*g} can be explicitly expressed as~\cite{AS3}
\begin{align}
\hat{t}_n^{(\eps k)} \ast \hat{t}_m^{(\eta l)}=
e^{\frac{i\hbar}{2}\theta^{ab}\partial_1^a\partial_2^b}
\hat{t}_n^{(\eps k)} \hat{t}_m^{(\eta l)}
=\exp{-\frac{i}{2}\hbar\omega (\eps kz\partial_2^z-\eta lz\partial_1^z)}
\hat{t}_n^{(\eps k)} \hat{t}_m^{(\eta l)}\,,
\end{align}
where $\omega$, related to equation \eqref{qbraA}, plays a crucial role as the {\it quantum dimension} as defined in \cite{AS3}. This parameter characterizes the magnitude of quantum space fluctuations and carries units of $\hbar$. Similarly, $k$ is termed the {\it quantum dimensional weight}, reflecting its function as the multiplier of $\omega$.

Let us introduce two fundamental operators: the scaling operator $\hat{S}_0$ and the normalized MT operator $\hat{T}_n^{(k)}$, defined respectively as
\beq
\hat{S}_0=q^{-2z\partial}\,,
\eeq
\beq
\Op{t}{m}{k} = z^m q^{-k(z\partial + \frac{m}{2} + \Delta)} =(\qqi)\Op{T}{m}{k}\,. \label{That} 
\eeq
Using these operators, we can express $\Op{T}{n}{k}$ as a product of $\Op{T}{n}{0}$ and a left-acting scaling operator $\hat{S}_0$:
\beq
\Op{T}{n}{k}=q^{k(\frac{n}{2}-\Delta)}\hat{S}_0^{\frac{k}{2}}\Op{T}{n}{0}\,.  \label{Tnk20}
\eeq

We define the $CZ^\pm$ generators in terms of MT operators~\cite{AS2,AS2e} according to
\beq
\hat{L}_n^\pm = \mp\Op{T}{n}{0} \pm q^{\pm(n+2\Delta)} \Op{T}{n}{\pm2}\,.  \label{Lpm1} 
\eeq
These generators satisfy a set of fundamental $\ast$-bracket commutation relations:
\begin{align}
&    [\Op{T}{n}{k},\Op{T}{m}{l}]_\ast =0\,,                 \label{*1}     \\
& [\hat{L}_n^\pm, \hat{L}_m^\pm]_\ast = [n-m]\hat{L}_{n+m}^\pm\,,   \label{*2}   \\
&  [\hat{L}_n^\pm,\Op{T}{m}{l}]_\ast = -[m]\Op{T}{n+m}{l}\,.  \label{*3}  
\end{align}
Furthermore, we obtain the $CZ^\ast$ algebra, which describes the interactions between $CZ^\eps$ generators ($\eps=\pm$), expressed as
\beq
 [L_n^\eps,L_m^\eta]_\ast =q^{\eta m}[n]L_{n+m}^\eps - q^{\eps n}[m]L_{n+m}^\eta\,,  \label{CZCZ}
\eeq

The general form of the $\ast$-bracket commutator is defined by
\beq
[X_n^{(k)},X_m^{(l)}]_\ast=(X_n^{(k)}X_m^{(l)})_\ast-(X_m^{(l)}X_n^{(k)})_\ast\,. \label{*4}
\eeq
This commutator applies to all elements $X_n^{\eps(k)}$ within the set $\mathscr{M}_T\{\hat{L}_n^\eps,T_n^{(k)}\}$ with weight $k$, where the $\ast$-product is defined as
\beq
(X_n^{\eps(k)}X_m^{\eta(l)})_\ast = q^{-x(\eps,\eta)} X_n^{\eps(k)} X_m^{\eta(l)}\,,
\quad\,\,  x(\eps,\eta)=\frac{\eta nl-\eps mk}{2} \,. \label{X*X}
\eeq
For the phase factor $x(\eps,\eta)$, we set $k=l=2$ when dealing with $\hat{L}_n^\eps$, while for $\hat{T}_n^{(k)}$ we use its intrinsic weight $k$. The signature parameter $\eps$ takes values $\pm$ for $\hat{L}_n^\pm$ operators, whereas it is fixed at $\eps=+$ for $\hat{T}_n^{(k)}$.

In Section~\ref{sec:*form}, we will examine the supersymmetric extension of the $CZ^+$ generator. For elements of the set 
$
X_n^{(k)} \in \mathscr{M}_S\{\mathcal{\hat{L}}_n,\mathcal{\hat{J}}_n, \mathcal{\hat{G}}_r\} 
$
the quantum dimensional weight $k$ exhibits a distinct pattern: it takes the value 2 for both $\mathcal{\hat{L}}_n$ and $\mathcal{\hat{J}}_n$ (where $\mathcal{\hat{J}}_n$ represents the $U(1)$ current corresponding to $\hat{T}_n^{(2)}$), while it equals 1 for the supercharge $\mathcal{\hat{G}}_r$. This distribution of weights reveals an underlying $Z_2$-grading structure in the quantum dimensional weight.

$CZ^\pm$ are the minimal subalgebras of the more general algebra $\mathcal{CZ}^\pm$, 
which are characterized by generators composed of the scaling operator $\hat{S}_0$ and $\hat{L}_n^\pm$ in the same way as \eqref{Tnk20}:
\beq
L_n^{\pm(k)}=q^{\mp k(\frac{n}{2}-\Delta)} \hat{S}_0^{\mp\frac{k}{2}} \hat{L}_n^\pm\,,
\label{Lnpmk}
\eeq
where
\beq
\hat{S}_0^\pm=1\pm (q-q^{-1})\hat{L}_0^\pm = q^{\mp2z\partial}\,.
\eeq
More generally, $L_n^{\pm(k)}$ satisfy both $\mathcal{CZ}^\ast$ and $\mathcal{CZ}^\pm$, with the $\ast$-bracket of $\mathcal{CZ}^\ast$ exhibiting a Moyal $\ast$-product structure~\cite{AS3}. Although $\hat{L}_n^\pm$ are derived by setting the weight $k=0$ in $L_n^{\pm(k)}$, the quantum dimensional weight should be set to 2 when reducing the $\ast$-brackets of $\mathcal{CZ}^\ast$ to those of $CZ^\ast$. 
Whether $L_n^{\pm(k)}$ can be redefined to reproduce $CZ^\ast$ at $k=2$ remains an open question beyond the scope of this study.

A significant property of the scaling operator $\hat{S}_0$ emerges when we assign its $k$th power $\hat{S}_0^k$ a weight of $2k$ in equation \eqref{X*X} for $k\not=0$. Under these conditions, $\hat{S}_0$ functions as a central element~\cite{AS}, satisfying
\beq
[\hat{S}_0^k, \hat{L}_n]_\ast =0\,,\quad [\hat{S}_0^k, \hat{T}_m^{(l)}]_\ast =0\,.
\eeq
Furthermore, we can express the $CZ^\pm$ generator \eqref{Lpm1} in an alternative form using the $q$-difference operator. By applying the normalization \eqref{That}, we obtain
\beq
\hat{L}_n^\pm = \mp z^{n}\frac{1-q^{\mp2z\partial}} {\qqi} 
=:-z^{n+1}\partial_q^\pm  \,.  \label{Ln_diff}
\eeq
This representation is equivalently referred to as either the $q$-difference or MT representation.

%
%%%%%%%%%%%%%%%%%%%%%%%%%%%%%%%%%%%%%%%%%%%%%%%%%%%%%%%%%%%%%%%%
%               Section 3   Scaled CZ algebra (L^\pm  algebra)
%
%%%%%%%%%%%%%%%%%%%%%%%%%%%%%%%%%%%%%%%%%%%%%%%%%%%%%%%%%%%%%%%%%
\setcounter{equation}{0}
\section{Supersymmetry and Quantum Superspace}\label{sec:QSS}
\indent

In this paper, we investigate a system where magnetic-spin interaction (Zeeman term) is introduced to induce supersymmetry in a system exhibiting CZ symmetry. For example, while a two-dimensional electron system under a static magnetic field exhibits 
CZ noncommutative structure (with quantum plane picture), our aim is to explore 
the relationship between super CZ emergence and quantum superspace (QSS) by extending this to a supersymmetric system.

As our approach, we apply SSM (superspace and spin matrix) correspondence to QSS 
to transform from QSS-based super CZ to MSB operator representation (Section~\ref{sec:MSB}). For rerefence, we organize SSM correspondence and super Virasoro algebra in the electron spin system under a static magnetic field 
in Appendix~\ref{sec:SV}. 
The key points are to remind the following two points while reviewing the setting of spin Grassmann basis in the electron spin system under static magnetic field: 
(i) Grassmann coordinate variables in superspace can be mapped to MSB (SSM correspondence). 
(ii) Using SSM correspondence, supercharge and Virasoro super generator based on superspace can be mapped to MSB representation with spin Grassmann basis.

To realize the super CZ algebras in MSB space, different CZ algebras emerge for the part acting on the up spin and the part acting on the down spin. In order to interpret this anomaly as an anomaly stemming from QSS, it is meaningful to systematically organize the super CZ algebras based on QSS formalism. Besides, in order to understand guideline of choosing which operator setting is suitable to consider, 
we examine the differences among three known types of super CZ algebras that emerge when replacing superspace with QSS in Section~\ref{sec:SCZ}. 
Type 1 represents the most straightforward approach, where the CZ algebra remains unmodified but a $U(1)$ current appears in the supercharge algebra. Type 2 features an anomaly in the CZ algebra while achieving partial simplification of the supercharge algebra. Type 3 exhibits a completely simplified supercharge algebra. 
\red{
A summary of the three types of CZ superalgebras is provided in Appendix~\ref{sec:3types} for the reader’s convenience.
}

Type 3 is unique in that its supercharge algebra can be expressed solely in terms of super Virasoro generators on the right-hand side, enabling its decomposition into $N=2$ supersymmetry. As we will demonstrate in Section~\ref{sec:MSB}, this Type 3 structure manifests naturally in electron spin systems, and exhibits the super $\ast$-bracket formalism presented in Section~\ref{sec:*form}.

%%%%%%%%%%%%%%%%%%%%%%%%%%%%%%%%%%%%%%%%%%%%%%
%  3.1    super CZ  (in QSS)  review
%%%%%%%%%%%%%%%%%%%%%%%%%%%%%%%%%%%%%%%%%%%%%%
\subsection{Super CZ algebras in quantum superspace (QSS)}\label{sec:SCZ}
\indent

Now we consider the case where the superspace $(x,\theta)$ is replaced with quantum superspace. Here we review three types of super CZ algebras in quantum superspace covariant under $GL_q(1,1)$ \cite{superCZ,superCZ2}. 
First, the bosonic fundamental operators are given by
\beq
B_n=-q^{-1}x^{n+1}\partial_x\,,\quad J_n=x^n\theta\partial_\theta\,. \label{Bn}
\eeq
These operators satisfy the deformed commutation relations:
\begin{align}
&[B_n,B_m]_{(m-n)}=[n-m]B_{n+m}\,,\quad [B_n,J_m]_{(m-n)}=-q^{-n}[m] J_{n+m}\,, \label{BBBFq} \\
& [J_n,J_m]_{(m-n)}=0\,.  \label{FFU1q}
\end{align}

The first super CZ algebra (Type 1) is constructed using the composite operator
\beq
 L_n=B_n-g_n J_n\,,\quad g_n=a q^{-2n}+b :=g_n^{CZ}\,,   \label{CZgen}
\eeq
where $L_n$ satisfies the same commutation relations as $B_n$ (we denote this representation as $CZ_{QSS}$, and $CZ_{QS}$ for the case when $g_n=0$):
\beq
[L_n,L_m]_{(m-n)}=[n-m]L_{n+m}\,,\quad[L_n,J_m]_{(m-n)}=-q^{-n}[m]J_{n+m}\,.
\label{type1LL}
\eeq
The supercharge is given by
\beq
G_r=\mu^{-\frac{1}{2}}x^{r+\frac{1}{2}}(\partial_\theta-\theta\partial_x)\,, \label{Gtype1}
\eeq
\beq
\mu=\partial_x x-x\partial_x=1+(q-q^{-1})L_0\,,  \label{mubyL0}
\eeq
and the anticommutation relation takes the form (after phase adjustment from Eq.(3.8) in \cite{superCZ2}):
\begin{align}
\{G_r,G_s\}_{(\frac{s-r}{2})}&=q^{r+s+2}(q^{\frac{s-r}{2}}+q^{\frac{r-s}{2}})B_{r+s}
-q^{\frac{r+s+3}{2}}([s+\frac{1}{2}]+[r+\frac{1}{2}])J_{r+s}\,,  \label{GrGs1} \\
&=q^{r+s+2}(q^{\frac{s-r}{2}}+q^{\frac{r-s}{2}})L_{r+s} \nn\\
&+\{ q^{2}(q^{\frac{3s+r}{2}}+q^{\frac{3r+s}{2}})g_{r+s}
-q^{\frac{r+s+3}{2}}([s+\frac{1}{2}]+[r+\frac{1}{2}]) \}J_{r+s}\,,  \label{GGtab1}
\end{align}
or alternatively
\beq
\{G_r,G_s\}_{(s-r)}=q^{r+s+2}(q^{s-r}+q^{r-s})B_{r+s}
-q^{\frac{3}{2}}(q^s[s+\frac{1}{2}]+q^r[r+\frac{1}{2}])J_{r+s}\,. \label{type1GG}
\eeq
If we choose
\beq
g_n=\frac{1}{[2]}q^{-n}[n+1]\,, \label{gn_1}
\eeq
then \cite{superCZ2}
\begin{align}
&[L_n,G_r]_{(r+\frac{1}{2}-n)}=q^{-n}[n-r-\frac{1}{2}]G_{n+r}+f_{n,r}\mu^{-n}G_{n+r}\mu^{n+1}\,,\label{LnGr1} \\
&f_{n,r}=q^{n-r-\frac{1}{2}-2n(n+r+\frac{3}{2})}\frac{[1-n]}{[2]}\,.
\end{align}
There are two issues with this form: first, the right-hand side of \eqref{GrGs1} is not expressed purely in terms of $L_{r+s}$, and second, the right-hand side of \eqref{LnGr1} has a complicated structure (the phase factor in the bracket is also somewhat peculiar and may need reconsideration - it would typically be $r-\frac{n}{2}$).

The remaining two types of super CZ algebra employ a slightly different composite form of $L'_n$:
\beq
 L'_n=B_n-g'_n J_n\,,\quad g'_n=a' q^{-n}+b'\,,  \label{CZTgen}
\eeq
The commutation relations for $L'_n$ receive the following modification (which we denote as $\,\scz$):
\begin{align}
&[L'_n,L'_m]_{(m-n)}=[n-m]L'_{n+m}+a_{n,m}J_{n+m}\,, \label{CZtwist} \\
&[L'_n,J_m]_{(m-n)}=-q^{-n}[m]J_{n+m}\,,
\end{align}
where
\beq
a_{n,m}=a'q^{-n-m}([m-n]+[n]-[m])=a'c^2q^{-n-m}
[\frac{n-m}{2}][\frac{n}{2}][\frac{m}{2}] \,,  \label{a_nm}
\eeq
\beq
c=\qqi\,.
\eeq

The super $\,\scz$ algebra allows several possibilities depending on the choice of $G_r$ and constants in $g'_n$, but only two specific forms have been studied in detail:
\begin{enumerate}
\item
Type 2 combination from \cite{superCZ2}:
\beq
g'_n=q^{-\frac{n}{2}}[\frac{n}{2}]\,\quad \mbox{with}\quad
G_r=\mu^{-\frac{1}{2}}x^r(\partial_\theta-x\theta\partial_x)\,, \label{Gtype2}
\eeq
\item
Type 3 combination from \cite{superCZ}:
\beq
g'_n=q^{-\frac{n+1}{2}}[\frac{n+1}{2}]\,,\quad  \mbox{with}\quad G_r
\,\,\mbox{given by}\,\eqref{Gtype1}   \label{Gtype3}
\eeq
\end{enumerate}
where $\mu$ is replaced by $\lambda$
\beq
\lambda=1+(q-q^{-1})L'_0\,.   \label{lambda}
\eeq
Note that in \eqref{Gtype2}, we have $\lambda=\mu$ since $g'_0=0$. For both Type 2 and Type 3 we have
\beq
[L'_n,G_r]_{(r-\frac{n}{2})}=q^{-n}[\frac{n}{2}-r]G_{n+r}\,. \label{LnGr2}
\eeq
However, the supercharge anticommutation relations take a simple closed form only for Type 3:
\beq
\{G_r,G_s\}_{(\frac{s-r}{2})}=q^{r+s+\frac{5}{2}}(q^{\frac{s-r}{2}}+q^{\frac{r-s}{2}})L'_{r+s}\,.
\label{GrGs2}
\eeq
For Type 2, the relations are similar to those of the first super CZ algebra \eqref{GrGs1}:
\beq
\{G_r,G_s\}_{(\frac{s-r}{2})}=q^{r+s+2}(q^{\frac{s-r}{2}}+q^{\frac{r-s}{2}})B_{r+s}
-q^{\frac{r+s+4}{2}}([s]+[r])J_{r+s}\,.  
\eeq
While this could be expressed in terms of $L'_{r+s}$ and $J_{r+s}$, we leave it in the formal expression due to its complexity (it is possible to find $b_{r,s}$ explicitly if necessary):
\beq
\{G_r,G_s\}_{(\frac{s-r}{2})}=q^{r+s+2}(q^{\frac{s-r}{2}}+q^{\frac{r-s}{2}})L'_{r+s}
+b_{r,s}J_{r+s}\,.  \label{GGtab2}
\eeq

A distinctive feature of Type 3 super $\,\scz$ algebra \eqref{LnGr2}, \eqref{GrGs2} is its natural decomposition into $N=2$ supersymmetry \cite{superCZ}. The supercharge decomposes as:
\beq
G_r=G_r^++G_r^-    \label{N2Gr}
\eeq
\beq
G_r^-=\lambda^{-\frac{1}{2}}x^{r+\frac{1}{2}}\partial_\theta\,,\quad
G_r^+=-\lambda^{-\frac{1}{2}}x^{r+\frac{1}{2}}\theta\partial_x\,,  \label{N2Grpm}
\eeq
\beq
\{G_r^+,G_s^-\}=q^{r+s+\frac{5}{2}}L'_{r+s}+q^{\frac{r-s+3}{2}}[\frac{r-s}{2}]J_{r+s}\,\,,\quad
\{G_r^\pm,G_s^\pm\}=0\,,  \label{N2GG}
\eeq
\beq
[L'_n,G_r^\pm]_{(r-\frac{n}{2})}=q^{-n}[\frac{n}{2}-r]G_{n+r}^\pm\,, \label{N2LG}
\eeq
\begin{align}
&[J_n,G_r^+]_{(\alpha,\beta)}=q^{n+2+\alpha} \lambda G_{n+r}^+ \,, \label{N2JG+}\\
&[J_n,G_r^-]_{(\alpha,\beta)}=-q^{n+2r+1+\beta} \lambda G_{n+r}^-\,, \label{N2JG-}
\end{align}
We do not yet know how to fix $\alpha$ and $\beta$. As we will see later, these free values will be determined when we impose physical requirements, for example, when we require them to correspond to the super CZ algebra of the electron spin system.

To conclude this section, we summarize the differences among the three types in the following table for comparison.
\captionsetup{width=\linewidth}
\begin{table}[H]
\centering
\begin{tabular}{|l|c|c|c|}
\hline \textbf{Property} & \textbf{Type 1} & \textbf{Type 2} & \textbf{Type 3} \\ \hline
%\hline & Type 1 & Type 2 & Type 3 \\ \hline
$g_n$ & $\frac{1}{[2]}q^{-n}[n+1]$ & $q^{-\frac{n}{2}}[\frac{n}{2}]$ & $q^{-\frac{n+1}{2}}[\frac{n+1}{2}]$ \\ \hline
$G_r$ form & $\mu^{-\frac{1}{2}}x^{r+\frac{1}{2}}(\partial_\theta-\theta\partial_x)$ & $\mu^{-\frac{1}{2}}x^r(\partial_\theta-x\theta\partial_x)$ & $\lambda^{-\frac{1}{2}}x^{r+\frac{1}{2}}(\partial_\theta-\theta\partial_x)$ \\ \hline
$[L_n,L_m]_{(m-n)}$  & $CZ$ & $\,\scz$ & $\,\scz$ \\ \hline
$[L_n,G_r]_{(\cdot)}$  & complex \eqref{LnGr1}& simple \eqref{LnGr2}& simple \eqref{LnGr2}\\ \hline
$\{G_r,G_s\}_{(\frac{s-r}{2})}$ & complex \eqref{GGtab1} & complex \eqref{GGtab2} & simple \eqref{GrGs2}\\ \hline
%$N=2$ decomp. & No & No & Yes \\ \hline
\end{tabular}
\caption{Structural comparison of three types of super CZ algebras in quantum superspace. For Types 2 and 3, the operators denoted by $L_n$ are in fact defined as the modified generators $L'_n$.}
\label{tab:compare}
\end{table}

%%%%%%%%%%%%%%%%%%%%%%%%%%%%%%%%%%%%%%%%
%     4.      super CZ in electron with Zeeman spin 
%%%%%%%%%%%%%%%%%%%%%%%%%%%%%%%%%%%%%%%%
\setcounter{equation}{0}
\section{QSS Correspondence: Mapping to Spin Matrix Space}\label{sec:MSB}
\indent

In this section, we develop a framework for realizing super CZ algebra on quantum superspace (QSS) through the magnetic-spin matrix basis (MSB) representation. This extends the existing correspondence between quantum space (QS) and magnetic translations (MT). To differentiate between the two realizations, we denote the CZ algebra constructed through MT operators as $CZ_{MT}$ and that constructed in quantum superspace as $CZ_{QSS}$.

The fundamental challenge in this construction lies in establishing the precise mapping of QSS to spin Grassmann basis within the $GL_q(1,1)$ framework. Although we can map QSS Grassmann operators $(\theta,\partial_\theta)$ to $(\sigma_1,\sigma_2)$ for construction in spin matrix (SM) space, we must first resolve how to properly map the bosonic operators $(x,\partial_x)$ through the QS-MT correspondence.

In Section~\ref{sec:mix}, we address a more subtle challenge: the realization of super $CZ_{MT}$ through application of the non-supersymmetric QS-MT correspondence to $CZ_{QSS}$. This investigation necessitates precise analysis of how the QSS scaling operator $\mu$ manifests within the spin Grassmann basis. A significant complexity arises from the fact that, unlike the non-super (QS) case, QSS possesses dual equivalent representations - bosonic and fermionic. We propose a novel operator mixing hypothesis to reconcile this QSS duality with its SM counterpart, establishing a coherent correspondence between QSS and SM representations.

Building upon this mixing hypothesis, we systematically derive the Type 3 super $\,\scz$ algebra in its MSB representation in Section~\ref{sec:proof}. Since the computational outline was already explained in the previous paper~\cite{SCZ4}, this section describes the computational details that were not presented there, and discusses the role of 
$\,\scz$ as a quantum effect from the mixing mechanism. 
Later in Section~\ref{sec:*form}, we demonstrate that both $N=1$ and $N=2$ super $\,\scz$ algebras can be formulated within a unified $\ast$-bracket framework, revealing their fundamental algebraic structures.

As the first step, we show that the CZ algebra in quantum space ($CZ_{QS}$) can be transformed into the magnetic translation CZ algebra ($CZ_{MT}$) by utilizing the correspondence between the QS differential operator set $(x,\partial_x,\mu)$ and the $q$-differential operator set $(z,\partial_q,\hat{\mu})$ (QS-MT correspondence).

Let us consider the correspondence between the QSS bosonic fundamental operator $B_n$ in \eqref{Bn} and the $q$-differential representation \eqref{Ln_diff} of $L_n$:
\beq
B_n=-q^{-1}x^{n+1}\partial_x\quad \leftrightarrow \quad \hat{L}_n=-z^{n+1}\partial_q\,.
\label{B2Lhat}
\eeq
From the formula
%which can be verified from \eqref{delfg}\eqref{delzn}
\beq
q\partial_q z=1+q^{-1}z\partial_q\,,\quad 
\eeq
and the bosonic differential part of the $GL_q(1,1)$ quantum plane
\beq
\partial_x x=1+q^{-2}x\partial_x\,,\quad 
\eeq
we recognize that the correspondence \eqref{B2Lhat} implies
\beq
(x,\partial_x) \quad \leftrightarrow \quad (z, q\partial_q)\,. \label{qpl2qdif}
\eeq

To verify the consistency of this correspondence, let us check other relations. For example:
\beq
\partial_x x^n=q^{-n+1}[n]x^{n-1} \quad\leftrightarrow \quad
q\partial_q z^n=q^{-n+1}[n]z^{n-1}
\eeq
\beq
\mu=\partial_x x-x\partial_x=1+(q-q^{-1})B_0 \quad\leftrightarrow \quad
\hat{\mu}=q(\partial_q z-z\partial_q)=1+(\qqi)\hat{L}_0 \label{mu2mu}
\eeq
\beq
\mu x^n \mu^{-1}=q^{-2n} x^n \quad\leftrightarrow \quad
\hat{\mu} z^n \hat{\mu}^{-1}=q^{-2n} z^n\,\quad\quad \mbox{\it{etc}.} \label{mu_rule}
\eeq
This establishes that quantum space differential operators correspond to $q$-differential operators, and since $q$-differential operators $\hat{L}_n$ correspond to MT operators $\Op{T}{n}{k}$, we conclude that differential operators in $GL_q(1,1)$ quantum space can be transformed into representations of MT operators.

Specifically, from \eqref{Ln_diff}, \eqref{That} and \eqref{Lpm1}, $\hat{L}_n$ can be decomposed as:
\begin{align}
&\hat{L}_n=\hat{B}_n+\hat{J}_n  \label{LbyBJ}\\
&\hat{B}_n=-\Op{T}{n}{0}=\frac{-z^n}{\qqi}   \label{BTz} \\
&\hat{J}_n=q^{n+2\Delta}\Op{T}{n}{2}=z^n\frac{q^{-2z\partial}}{\qqi}\,. \label{JTz}
\end{align}
The commutation relations derived from
\beq
[\Op{T}{n}{k}, \Op{T}{m}{l}]_{(m-n)}=[\frac{n(l-2)-m(k-2)}{2}]\Op{T}{n+m}{k+l}\,,
\eeq
yield 
\begin{align}
&[\hat{B}_n,\hat{B}_m]_{(m-n)}=[n-m]\hat{B}_{n+m}\,,\\
& [\hat{B}_n,\hat{J}_m]_{(m-n)}=-q^{-n}[m] \hat{J}_{n+m}\,, \quad
 [\hat{J}_n,\hat{J}_m]_{(m-n)}=0\,.  \label{hatBJJJ}
\end{align}
These are isomorphic to $CZ_{QS}$, namely to \eqref{BBBFq} and \eqref{FFU1q}. 

Following the composition rule \eqref{CZgen}, the CZ operator is given by
\beq
\hat{L}_n^{CZ}=\hat{B}_n - g_n^{CZ}\hat{J}_n   \label{CZgen2}
\eeq
and satisfies the CZ algebra (which we denote as $CZ_{MT}$). Similarly, replacing $g_n^{CZ}$ with $g'_n$ yields the $\,\scz$ algebra. Note that \eqref{LbyBJ} corresponds to the special case of these ($a=0,b=-1$).

From \eqref{qpl2qdif}, the relationship represents QS-MT correspondence, but while $(B_n,J_n)$ and $(\hat{B}_n,\hat{J}_n)$ play similar roles as constituent elements of CZ operators, it is clear from \eqref{BTz} and \eqref{JTz} that they do not coincide in the $q\to 1$ limit. (From another viewpoint, the counterpart of $B_n$ is not $\hat{B}_n$ but $\hat{L}_n$.) Hence, we cannot simply apply this correspondence, and there must be a mixing matrix transformation between $(B_n,J_n)$ and $(\hat{B}_n,\hat{J}_n)$ (as will be discussed later in Section~\ref{sec:mix}). As a result, the algebra in the $q\to 1$ limit may differ from the one expected from the QSS viewpoint. 
Therefore, assuming this mixing, we aim to construct the superalgebra in the QSS framework for $q \neq 1$, while addressing a specific scenario where the supercharge becomes $q$-anticommutative as suggested in \cite{super3,HHT} when $q \to 1$.

Before proceeding further, let us note the MT counterpart $\hat{\mu}$ of the QS scaling operator $\mu$. From \eqref{Ln_diff} and \eqref{qpl2qdif}, we have
\beq
\hat{\mu}=1-(\qqi)z\partial_q=q^{-2z\partial}\,.  \label{muhat}
\eeq
This gives the transmutation rule between $\hat{J}_n$ and $\hat{B}_n$ as
\beq
\hat{J}_n=\frac{z^n\hat{\mu}}{\qqi}=-\hat{B}_n\hat{\mu}\,,  \label{JBmu}
\eeq
and making use of \eqref{mu_rule} we verify the scaling rule 
\beq
\hat{\mu}\hat{X}_n\hat{\mu}^{-1}=q^{-2n}\hat{X}_n\,,\quad 
\hat{X}_n=\hat{L}_n,\hat{B}_n,\hat{J}_n\,.
\eeq
Note that all $\Op{T}{n}{k}$ can be separated into products of $\hat{B}_n$ and $\hat{\mu}$:
\beq
\Op{T}{n}{k}=q^{-k(\frac{n}{2}+\Delta)}\Op{T}{n}{0} \hat{\mu}^{\frac{k}{2}}\,, \label{Tnk2}
\eeq
\beq
\Op{J}{n}{k}:=q^{k(n+2\Delta)}\Op{T}{n}{2k}=-\hat{B}_n\hat{\mu}^k\,.
\eeq

The correspondence \eqref{mu2mu} represents the non-super pure bosonic correspondence. In the super CZ case with QSS, we originally have from \eqref{mubyL0}:
\beq
\mu=1+(\qqi)L_0=1+(\qqi)(B_0-g^{CZ}_0 J_0)\,, \label{mu_super}
\eeq
which holds under the pure bosonic condition ($g^{CZ}_0=0$). Similarly for:
\beq
\lambda=1+(\qqi)L_0'=1+(\qqi)(B_0-g'_0 J_0)\,. \label{lambda_super}
\eeq

When considering the MSB representation of super CZ, we need to consider both bosonic (MT) and Grassmann (SM) representations. In $GL_q(1,1)$ quantum superspace, we have:
\beq
\mu=\partial_x x - x\partial_x=\partial_\theta\theta+\theta\partial_\theta \label{muxt}
\eeq
This indicates that while the MT representation of $\mu$ corresponds to the scaling operation on $z$ as $\hat{\mu}=q^{-2z\partial}$, its SM representation must map to the unit matrix under the replacement correspondence \eqref{grass}:
\beq
\mu \mapsto \sigma_1\sigma_2+\sigma_2\sigma_1=\mathbbm{1}\,. \label{mumu1}
\eeq
It should also be noted that considering $\mu$ is relevant for Type 1 and/or Type 2, while for Type 3, $\lambda$ rather than $\mu$ serves as the scaling operator. The relation between $\lambda$ and $\mu$ is:
\beq
\lambda=\mu-(1-q^{-1})\theta\partial_\theta
=\partial_\theta\theta+q^{-1}\theta\partial_\theta
\eeq
Compared to \eqref{muxt}, the coefficient of $\theta\partial_\theta$ becomes $q^{-1}$, thus the SM representation is:
\beq
\lambda \ra q^{-1}\sigma_1\sigma_2+\sigma_2\sigma_1=diag(1,q^{-1})\,, \label{lambda_mat}
%=\begin{pmatrix}
%1 & 0  \\
%0 & q^{-1}  \\
%\end{pmatrix}
\eeq
which introduces a phase distortion in the down-spin component compared to \eqref{mumu1}.

Returning to the differential operator (MT) representation, recall that the QSS scaling operators $\mu$ or $\lambda$ are given by \eqref{mu_super} and \eqref{lambda_super}, where the supersymmetric case operators $L_0$ and $L'_0$ are extensions of the pure bosonic $B_0$. While $B_0$ does not directly correspond to $\hat{B}_0$, if we tentatively assume such a correspondence and consider the MT mapping
\beq
(B_0, J_0) \quad \ra \quad (\hat{B}_0, \hat{J}_0)\,,
\eeq
we obtain
\beq
\mu \,\ra\, \hat{B}_0-g^{CZ}_0 \hat{J}_0=\hat{B}_0-(a+b) \hat{J}_0=-(a+b)\hat{\mu}\,,
\eeq
where $\hat{\mu}$ is the same $q^{-2z\partial}$ as the bosonic $\mu$ counterpart.
Here, $(a,b)$ are defined in \eqref{CZgen}. Similarly for $\lambda$ (replacing $g^{CZ}_0$ with $g'_0$) we have
\beq
\lambda\,\ra\, \hat{B}_0-g'_0 \hat{J}_0=-(a'+b')\hat{\mu}\,, \label{scale2}
\eeq
where $(a',b')$ are defined in \eqref{CZTgen}.

The values of $(a,b)$ remain undetermined, as we cannot fix them at this stage. This is because there is no guarantee that the $g^{CZ}_n$ in \eqref{CZgen2} is identical to that in QSS. For example, if we assume the correspondence
\beq
(B_0, J_0) \quad \ra \quad (\hat{B}_0+\beta \hat{J}_0, \hat{J}_0)\,, %+\alpha \hat{B}_0)
\label{BJmix}
\eeq
the effect of $\beta$ can be absorbed into the coupling coefficients of the MT representation by taking
\beq
g_n^{CZ}=aq^{-2n}+ b+\beta\,,\quad g'_n=a'q^{-2n}+ b'+\beta
\eeq
to maintain the same results.
%$\alpha$の効果は$a+b=0$と選べば消すことができて$\mu,\lambda\ra\beta\hat{\mu}$となる。

%\vskip\baselineskip
%%%%%%%%%%%%%%%%%%%%%%%%%%%%%%%%%%
%  4.1    QSS correspondence
%%%%%%%%%%%%%%%%%%%%%%%%%%%%%%%%%%
\subsection{Mixing transformation}\label{sec:mix}
\indent

When applying the SSM correspondence \eqref{grass} to the supercharge \eqref{Gtype1}, we observe that $B_n$ and $F_n (=x^n)$ emerge in the transformed expression. Rather than directly applying our previous analysis, we must first establish a precise correspondence for $F_n$, as this forms the foundation for understanding the dual nature of the QSS scaling operator.

To establish the mathematical framework, we first systematically organize the dual representation of the QSS scaling operator $\mu$, drawing from the relations established in \eqref{mu2mu}, \eqref{muxt}, and \eqref{mumu1}, as shown in Table~\ref{tb:mu}.
\begin{table}[H]
\centering
\begin{adjustbox}{max width=\linewidth}
\begin{tabular}{|l|c|c|} \hline
   coord. system & QSS rep. $(\mu)$ & MSB rep. \\ \hline
   bosonic & $\partial x-x\partial$ & $\hat{\mu}$  \\ \hline
   grassmann  & $\theta\partial_\theta+\partial_\theta\theta$ & $\mathbbm{1}$   \\ \hline
 \end{tabular}
\end{adjustbox}
\caption{Dual representation of QSS scaling operator $\mu$ and its corresponding operators. }
\label{tb:mu}
\end{table}

From the bosonic quantum space $q$-differential (MT operator) correspondence \eqref{qpl2qdif} and \eqref{BTz}, we have $F_n=x^n\mapsto z^n=-c \hat{B}_n,\, (c=\qqi)$. While the scaling operation part $\theta\partial_\theta$ isolated from $J_n$ can be rewritten in terms of $\mu$, resulting in a bosonic scaling operator, it leaves a residual Grassmann operator term ($\mu-\partial_\theta\theta$). If we consider a twisted $J_n$ with $x^n\partial_\theta\theta$ added:
\beq
\mathfrak{J}_n=J_n+x^n\partial_\theta\theta=F_n\mu \,,  \label{mixJ}
\eeq
we then obtain a generalized correspondence relation in Table~\ref{tb:u1} that includes the $\mu$ case of Table~\ref{tb:mu} at $n=0$, since $\mathfrak{J}_0=\mu$. The factor $-c$ obtained from the $F_n$ part has been removed.
\begin{table}[H]
\centering
\begin{adjustbox}{max width=\linewidth}
\begin{tabular}{|l|c|c|} \hline
   coord. system & QSS rep. $(\mathfrak{J}_n)$ & MSB rep. \\ \hline
   bosonic & $x^n(\partial x-x\partial)$ & $\hat{B}_n\hat{\mu}=-\hat{J}_n$  \\ \hline
   grassmann  & $x^n(\theta\partial_\theta+\partial_\theta\theta)$ & $\hat{B}_n\mathbbm{1}$   \\ \hline
 \end{tabular}
\end{adjustbox}
\caption{Dual representation of QSS twisted operator $\mathfrak{J}_n$ and its potential MSB correspondence. }
\label{tb:u1}
\end{table}
This analysis reveals that the fundamental correspondence $\mu\mapsto(\hat{\mu},\mathbbm{1})$ from QSS to MSB naturally extends to the more general mapping $F_n\mapsto(\hat{J}_n,\hat{B}_n)$. 
Furthermore, the relationship between $L_n$ and $\hat{L}_n$ admits a more general mixing structure than \eqref{BJmix}, arising from two key observations: first, the coupling coefficients $g_n^{CZ}$ in the $CZ_{QSS}$ operator \eqref{CZgen} and $CZ_{MT}$ operator \eqref{CZgen2} need not coincide; second, \eqref{CZgen2} does not require the pure bosonic condition ($g_n^{CZ}=0$). This leads to the general mixing transformation:
\beq
\begin{pmatrix}
{B}_n  \\
{J}_n  \\
\end{pmatrix} 
\leftrightarrow
\begin{pmatrix}
m_{11} &  m_{12}  \\
m_{21} & m_{22}  \\
\end{pmatrix} 
\begin{pmatrix}
\hat{B}_n  \\
\hat{J}_n  \\
\end{pmatrix} \,.
\eeq

To constrain our general mixing transformation while preserving essential physical features, we impose two key conditions. First, to maintain consistency with the non-supersymmetric case, we require the bosonic part to exhibit proportional behavior. Second, to avoid over-generalization, we set $m_{12}=0$. These constraints lead to a simplified form for the MSB counterparts $(\tilde{B}_n,\tilde{J}_n)$ of $(B_n, F_n)$:
\beq
\tilde{B}_n:=m_{11}\hat{B}_n\,,\quad
\tilde{J}_n:=m_{21}\hat{B}_n+m_{22}\hat{J}_n \,. \label{JtildeBJ}
\eeq
Although a possible rotational symmetry might further reduce the degrees of freedom, this representation proves sufficient for realizing our target super algebra, as demonstrated in subsequent sections.

Substituting $(\tilde{B}_n,\tilde{J}_n)$ for the $(B_n,F_n)$ components in supercharge \eqref{Gtype1}, we obtain a candidate for the MSB counterpart of the supercharge:
\beq
\hat{\mathcal{G}}_r=\hat{\mu}^{-\frac{1}{2}} 
\tilde{B}_{r-\frac{1}{2}}\otimes\sigma_1+
\hat{\mu}^{-\frac{1}{2}}\tilde{J}_{r+\frac{1}{2}}\otimes\sigma_2\,.  \label{hatGr}
\eeq
This expression provides a foundation for analyzing the $q$-anticommutator $\{\mathcal{\hat{G}}_r,\mathcal{\hat{G}}_s\}_{(\frac{s-r}{2})}$. A crucial observation emerges when we expand this $q$-anticommutator in terms of $\hat{B}_{r+s}$ polynomials: the relations \eqref{CZgen2} and \eqref{JBmu} require that these terms combine to form $L_n^{CZ}$ with both zeroth and first order terms in $\hat{\mu}$. This requirement imposes a significant constraint on the mixing parameters: $m_{21}m_{22}\neq 0$. Indeed, if $\tilde{J}_n$ were constructed using only $\hat{B}_n$ or $\hat{J_n}$ ($m_{21}m_{22}=0$), we would obtain terms of a single order in $\hat{\mu}$, making it impossible to achieve the required structure.

A detailed analysis of the $q$-anticommutator reveals terms of two distinct orders: $\mathcal{O}(\hat{\mu}^{-1})$ and $\mathcal{O}(1)$. This structure suggests that, after appropriate coefficient adjustments, the complete expression should assume the form $\hat{\mu}^{-1}L^{CZ}_{r+s}$. However, the presence of the $\hat{\mu}^{-1}$ factor necessitates a modification of our mixing matrix. Specifically, $\tilde{J}_n$ must incorporate an additional $\hat{\mu}$ factor, leading to the refined mixing matrix:
\beq
\begin{pmatrix}
m_{11} &  0  \\
m_{21}\hat{\mu} & m_{22}\hat{\mu}  \\
\end{pmatrix} \,.
\eeq
Incorporating the $\hat{\mu}$ dependence while preserving the fundamental structure of \eqref{JtildeBJ}, we obtain a generalized form of the supercharge \eqref{hatGr}:
\beq
\hat{\mathcal{G}}_r=\hat{\mu}^{-\nu} 
\tilde{B}_{r-\frac{1}{2}}\otimes\sigma_1+
\hat{\mu}^{\nu}\tilde{J}_{r+\frac{1}{2}}\otimes\sigma_2
\quad\mbox{with}\quad\nu=\frac{1}{2} \,.    \label{hatGr2}
\eeq

As we will see in Section~\ref{sec:proof}, the power of $\hat{\mu}$ is not restricted to $\nu=\frac{1}{2}$ but can be generalized with $\nu$ left undetermined. With this generalization, we can reproduce the Type 3 $GL_q(1,1)$ super $\,\scz$ algebra \eqref{CZtwist}, \eqref{LnGr2} and \eqref{GrGs2}. Moreover, in the case of $\nu=1$, we can reproduce the Type 3 $GL_q(1,1)$ $N=2$ super $\,\scz$ algebra \eqref{N2GG} etc. 
There is no correspondence to Type 2 super $\,\scz$.

%\vskip\baselineskip
%%%%%%%%%%%%%%%%%%%%%%%%%%%%%%%%%%%%%%%%
%  4.2   verification
%%%%%%%%%%%%%%%%%%%%%%%%%%%%%%%%%%%%%%%%
\subsection{Remarks on verification}\label{sec:proof}
\indent

This section is devoted to explaining several remarks that were not addressed in the previous paper~\cite{SCZ4}. 
Building upon the theoretical framework established in the previous section, we now introduce a general parameterization of $\hat{\mathcal{G}}_r$ to determine its precise form through algebraic consistency requirements:
\beq
\mathcal{\hat{G}}_r=\hat{\mu}^{-\nu} \tilde{B}_r \otimes\sigma_1+c
\hat{\mu}^{\nu} \tilde{J}_r \otimes\sigma_2\,, \label{Gform}
\eeq
\beq
\tilde{B}_r= q^\gamma \hat{B}_{r-\frac{1}{2}}\,,\quad
\tilde{J}_r=-(q^\alpha \hat{J}_{r+\frac{1}{2}} + q^\beta \hat{B}_{r+\frac{1}{2}})\,.
\label{BJtilde}
\eeq
The construction of $q$-anticommutator requires careful treatment of the $\hat{\mu}$ powers in its cross terms. We configure these terms with inverse powers to achieve mutual cancellation. The structure of $L^{CZ}_{r+s}$ naturally emerges from two fundamental types of products: $\hat{J}_{r+\frac{1}{2}}\hat{B}_{s-\frac{1}{2}}\sim\hat{J}_{r+s}$ and 
$\hat{B}_{r+\frac{1}{2}}\hat{B}_{s-\frac{1}{2}}\sim\hat{B}_{r+s}$. This observation motivates our construction of $\tilde{J}_r$ as a linear combination of $\hat{J}_{r+\frac{1}{2}}$ and $\hat{B}_{r+\frac{1}{2}}$. The coupling constants are chosen proportional to $c$, ensuring the emergence of pure bosonic behavior in the classical limit $q\to 1$. This formulation reveals super $\,\scz$ as a quantum effect arising from the mixing matrix structure. The normalization factor $c=\qqi$ can be absorbed into the generators $\mathcal{\hat{G}}_r$ and $\mathcal{\hat{J}}_n$, simplifying the final expressions. 
This factor originates naturally from the normalization in the fusion rule \eqref{Trans}, and could alternatively be eliminated through operator redefinition as demonstrated in \eqref{trans}. The proportionality of the anomaly coefficient $a_{n,m}$ to $c^2$ in the second term of super $\,\scz$ provides additional theoretical justification: the anomaly's vanishing behavior as $q\to 1$ aligns with our physical expectations for the classical limit.

To proceed with the verification, we introduce a key set of assumptions regarding parameter dependencies: $\nu$ remains independent of $r$, while the parameters 
$\alpha$, $\beta$, and $\gamma$ are allowed to have $r$-dependence:
\beq
\alpha=\alpha(r)\,,\quad\beta=\beta(r)\,,\quad \gamma=\gamma(r)\,. \label{abc}
\eeq
For expressions involving $\mathcal{\hat{G}}_s$, we adopt the conventional notation 
$\alpha'=\alpha(s)$, with analogous expressions for the other parameters. 
The explicit computations proceed through systematic application of the fusion rule \eqref{Trans}. After some lengthy calculations, one can derive Eq.(79) 
shown in~\cite{SCZ4} as factorization conditions:
\beq
(A_+,B_+)=(C_+, D_+)\,,\quad\mbox{and}\quad (A_-,B_-)=(D_-,C_-)\,.\label{ABCD}
\eeq
These conditions are relevant to the following system of equations:
\begin{align}
&\alpha+1+2r=\beta\,,\quad \alpha'+1+2s=\beta'\,, \label{eq1} \\
&(2\nu-1)(r-s)+1+r+s+ \gamma-\gamma'+\alpha'=\beta \,, \label{eq2}\\
&(2\nu-1)(s-r)+1+r+s+ \gamma'-\gamma+\alpha=\beta' \,. \label{eq3}
\end{align}
The general solution takes the form
\beq
\alpha(r)=-r-\frac{1}{2}+e_1\,,\quad \beta(r)=r+\frac{1}{2}+e_1\,,\quad \gamma(r)=(1-2\nu)r+e_1+e_2\,, \label{abc_def}
\eeq
where $e_1$, $e_2$ and $\nu$ are arbitrary parameters. Specifically, setting
\beq
e_1=0\,,\quad e_2=\nu+\frac{1}{2} \,,\label{e1e2_def}
\eeq
we obtain
\beq
\{\mathcal{\hat{G}}_r,\mathcal{\hat{G}}_s  \}_{(\frac{s-r}{2})}=c[\frac{r-s}{2}]_+q^{1+r+s}
\mathcal{\hat{L}}_{r+s}\,, \label{newGGL}
\eeq
\beq
\mathcal{\hat{L}}_n=\begin{pmatrix}
q^{-2n\nu}\mathcal{\hat{L}}_n^+ & 0  \\
0 & \mathcal{\hat{L}}_n^-  \\
\end{pmatrix} \,, \label{supLmat}
\eeq
where the $CZ$ and $\,\scz$ operators are defined as
\beq
\mathcal{\hat{L}}_n^+=\hat{B}_n+q^{-2n}\hat{J}_n\,,\quad
\mathcal{\hat{L}}_n^-=\hat{B}_n+q^{-1-n}\hat{J}_n\,, \label{supLpm}
\eeq
and
\beq
[\frac{r-s}{2}]_+=\frac{q^{\frac{r-s}{2}}+q^{\frac{s-r}{2}}}{\qqi}\,,
\eeq

The operators $\mathcal{\hat{L}}_n^\pm$ are characterized by the following commutation relations:
\beq
[\mathcal{\hat{L}}_n^+,\mathcal{\hat{L}}_m^+]_{(m-n)}=[n-m]\mathcal{\hat{L}}_{n+m}^+\,,\quad
[\mathcal{\hat{L}}_n^-,\mathcal{\hat{L}}_m^-]_{(m-n)}=[n-m]\mathcal{\hat{L}}_{n+m}^- +a_{n,m}\hat{J}_{n+m}\,. \label{CZCZ'}
\eeq
This demonstrates that $\mathcal{\hat{L}}_n$ satisfies the $\,\scz$ algebra with parameters $a'=q^{-1}$ and $b'=0$ (cf. \eqref{CZTgen} and \eqref{supLpm}). Consequently, we confirm:
\beq
[\mathcal{\hat{L}}_n,\mathcal{\hat{L}}_m]_{(m-n)}=[n-m]\mathcal{\hat{L}}_{n+m}
 +\frac{1}{c}a_{n,m}\mathcal{\hat{J}}_{n+m} \,,\label{LLtwist}
\eeq
\beq
[\mathcal{\hat{L}}_n,\mathcal{\hat{G}}_r]_{(r-\frac{n}{2})}=q^{-n}[\frac{n}{2}-r]\mathcal{\hat{G}}_{r+s}\,. \label{N1LG}
\eeq
The matrix representation of $\hat{J}_n$ is given as
\beq
\mathcal{\hat{J}}_n=c\hat{J}_{n}\otimes\sigma_1\sigma_2\,, \label{Jmat}
\eeq
and its commutation relations satisfy the same form as in \eqref{hatBJJJ}:
\begin{align}
& [\mathcal{\hat{L}}_n,\mathcal{\hat{J}}_m]_{(m-n)}=-q^{-n}[m] \mathcal{\hat{J}}_{n+m}\,,  \label{calLJJJ}  \\
& [\mathcal{\hat{J}}_n,\mathcal{\hat{J}}_m]_{(m-n)}=0\,.  
\end{align}
These results demonstrate that the MSB representation precisely realizes a Type 3 super $\,\scz$ algebra. Specifically, \eqref{LLtwist} and \eqref{N1LG} match the QSS representation in \eqref{CZtwist} and \eqref{LnGr2}, while \eqref{newGGL} corresponds to \eqref{GrGs2} up to an overall factor of $q^{3/2}$ that is absent from the right-hand side. This missing factor can be absorbed through a redefinition of $\mathcal{\hat{G}}_r$.

Finally, we put a remark on the algebraic structure of the $N=2$ commutation 
relations between $\mathcal{\hat{J}}_n$ and $\mathcal{\hat{G}}_r^\pm$:
\beq
[\mathcal{\hat{J}}_n,\mathcal{\hat{G}}_r^\pm]_{(p_2)}=
\pm q^{\pm p_2+n+r+\frac{1}{2}\mp(r+\frac{1}{2})}\hat{\mu}\mathcal{\hat{G}}_{n+r}^\pm\,.
\eeq
It is very tempting to choose $p_2=r+\frac{1}{2}$, 
however, that does not allow for values of $\alpha=-\beta$ that match the 
corresponding relations \eqref{N2JG+} and \eqref{N2JG-} in the QSS case.
To this end, we have to redefine $\mathcal{\hat{J}}_n$ using the scaling freedom as:
\beq
\mathcal{\hat{J}}_n^{'}=q^{w}\mathcal{\hat{J}}_n\,, \label{newJ}
\eeq
and choose $p_2=r-\frac{n}{2}$ and $w=w'=1$ with setting 
$\alpha=-\beta=r-\frac{n}{2}-1$ in \eqref{N2JG+} and \eqref{N2JG-}.

Through these calculations, we have successfully reproduced the $GL_q(1,1)$ super $\,\scz$ algebra \eqref{CZtwist}, \eqref{LnGr2}, \eqref{GrGs2} in the MSB representation, which is identical to that in QSS. Furthermore, for $\nu=1$, we have reproduced the Type 3 $GL_q(1,1)$ $N=2$ super $\,\scz$ algebra \eqref{N2GG} and related relations. Note that there is no correspondence with the Type 2 super $\,\scz$ algebra. 

We have obtained the super $\,\scz$ matrix $\mathcal{\hat{L}}_n$ where the $CZ$ operator $\mathcal{\hat{L}}_n^+$ acts on up spin states and the $\,\scz$ operator $\mathcal{\hat{L}}_n^-$ acts on down spin states. The super $\,\scz$ matrix is generated from the supercharges $\mathcal{\hat{G}}_r$ and $\mathcal{\hat{G}}_r^\pm$. In the $N=2$ case, $\mathcal{\hat{L}}_n$ exhibits a particularly simple structure: it can be constructed directly from anticommutators that do not involve the deformation parameter $q$, maintaining the classical form of the $N=2$ supersymmetry algebra.

Comparing with the $q=1$ case, we find both similarities and key differences in structure. The similarity lies in the appearance of $B_n$ and $L_n^B$ as diagonal components (see \eqref{BL2let}). However, a fundamental difference appears in the coupling coefficients of $B_n$ and $F_n$ when comparing \eqref{EYform1} or \eqref{EYform2} with \eqref{supLpm} and \eqref{supLmat}. Another distinctive feature emerges in the $F_n$ terms: while the $F_n$ terms in \eqref{Gform2}-\eqref{EYform2} are essentially replaced by $\hat{J}_n$, in the case of \eqref{Gform2}, they transform into $\tilde{J}_n$ - a mixed state of $B_n$ and $J_n$ (see \eqref{Gmat}).

%\vskip\baselineskip
%%%%%%%%%%%%%%%%%%%%%%%%%%%%%%
%     5    *-bracket formalism
%%%%%%%%%%%%%%%%%%%%%%%%%%%%%%
\setcounter{equation}{0}
\section{Supersymmetric $\ast$-Bracket Formalism}\label{sec:*form}
\indent

Our analysis in Sections~\ref{sec:mix} and \ref{sec:proof} demonstrates that the MSB representation faithfully reproduces the structure of Type 3 super $\,\scz$ algebra, establishing a complete isomorphism with the QSS formulation. This result validates the correspondence between quantum superspace and physical spin systems. Building upon this foundation, we now systematically organize these results within the framework of $\ast$-bracket formalism to reveal their underlying algebraic structure.

We examine the supersymmetric extension of the $\ast$-bracket \eqref{X*X} 
for the set
\beq
X_n^{(k)} \in \mathscr{M}_S\{\mathcal{\hat{L}}_n,\mathcal{\hat{J}}_n, \mathcal{\hat{G}}_r\} 
\label{superM}
\eeq
where we assume $\eps=\eta=1$ since our superalgebra is an extension of $CZ^+$. 
We show that the quantum dimensional weight $k$ exhibits a distinct pattern: it takes the value 2 for both $\mathcal{\hat{L}}_n$ and $\mathcal{\hat{J}}_n$, while it equals 1 for the supercharge $\mathcal{\hat{G}}_r$. This distribution of weights reveals an underlying $Z_2$-grading structure in the quantum dimensional weight. 

In the case of $N=2$, we present that a nontrivial structure appears 
in the $\ast$-brakets for the supercharges. The role of $\eps$ and $\eta$ 
revives in terms of the $N=2$ decomposition. 

%%%%%%%%%%%%%%%%%% %
%  5.1   N=1
%%%%%%%%%%%%%%%%%%%
\subsection{The $N=1$ case}\label{sec:N1*}
\indent

The MSB representation of $CZ$ generators given by \eqref{supLmat} 
and $U(1)$ current $\mathcal{\hat{J}}_n$ given by \eqref{Jmat}, after incorporating the redefinition, are:
\beq
\mathcal{\hat{L}}_n=q^{-2n\nu}\mathcal{\hat{L}}_n^+ \otimes\sigma_2\sigma_1
+\mathcal{\hat{L}}_n^- \otimes\sigma_1\sigma_2
\eeq
\beq
\mathcal{\hat{J}}_n=qc\hat{J}_{n}\otimes\sigma_1\sigma_2
\eeq
where $CZ$ and $\,\scz$ generators $\mathcal{\hat{L}}_n^\pm$ are given by \eqref{supLpm}.
These satisfy the following $N=1$ super $\,\scz$ algebra as shown in \eqref{newGGL},\eqref{LLtwist}, \eqref{N1LG} and \eqref{calLJJJ}:
\beq
[\mathcal{\hat{L}}_n,\mathcal{\hat{L}}_m]_{(m-n)}=[n-m]\mathcal{\hat{L}}_{n+m}
 +cq^{-1}\, \alpha_{n,m}\mathcal{\hat{J}}_{n+m} \,, \label{superLL}
\eeq
\beq
\{\mathcal{\hat{G}}_r,\mathcal{\hat{G}}_s  \}_{(\frac{s-r}{2})}=
(q^{\frac{r-s}{2}}+q^{\frac{s-r}{2}}) q^{1+r+s}\mathcal{\hat{L}}_{r+s}\,, 
\eeq
\beq
[\mathcal{\hat{L}}_n,\mathcal{\hat{G}}_r]_{(r-\frac{n}{2})}=q^{-n}[\frac{n}{2}-r]\mathcal{\hat{G}}_{r+s}\,. 
\eeq
\beq
 [\mathcal{\hat{L}}_n,\mathcal{\hat{J}}_m]_{(m-n)}=-q^{-n}[m] \mathcal{\hat{J}}_{n+m}\,,  \quad
 [\mathcal{\hat{J}}_n,\mathcal{\hat{J}}_m]_{(m-n)}=0\,, \label{calhatJJ}
\eeq
where the diagonal elements $\mathcal{\hat{L}}_n^\pm$ satisfy \eqref{CZCZ'}, namely the $CZ$ and $\,\scz$ algebras respectively:
\beq
[\mathcal{\hat{L}}_n^+,\mathcal{\hat{L}}_m^+]_{(m-n)}=[n-m]\mathcal{\hat{L}}_{n+m}^+\,,\quad
[\mathcal{\hat{L}}_n^-,\mathcal{\hat{L}}_m^-]_{(m-n)}=[n-m]\mathcal{\hat{L}}_{n+m}^- +c^2\,\alpha_{n,m} \hat{J}_{n+m}\,.   \label{sumLL}
\eeq
Here, $\alpha_{n,m}$ is related to $a_{n,m}$ of \eqref{a_nm} through $\alpha_{n,m}=c^{-2}a_{n,m}$ with $a'=q^{-1}$, giving:
\beq
\alpha_{n,m}= q^{-1} q^{-n-m}
[\frac{n-m}{2}][\frac{n}{2}][\frac{m}{2}] \,.
\eeq

All equations from \eqref{superLL} to \eqref{sumLL} can be unified within the $\ast$-bracket formalism for $\mathscr{M}_S$ introduced in \eqref{superM}. This formalism extends the commutator \eqref{*4} by incorporating $Z_2$-grading, while preserving the direct use of the $\ast$-bracket formula \eqref{X*X} with setting $\eps=\eta=1$. 
The quantum dimensional weights are assigned as $k,l=1$ for the supercharge and $k,l=2$ for other generators:
\beq
[X_n^{(k)},X_m^{(l)}]_\ast=(X_n^{(k)} X_m^{(l)})_\ast-(-1)^{deg(X_n^{(k)})deg(X_m^{(l)})}(X_m^{(l)} X_n^{(k)})_\ast \label{*5}
\eeq
where the grading function is defined as
\beq
deg(X_n^{(k)})=1 \quad\mbox{for}\quad\mathcal{\hat{G}}_r\,,\quad\mbox{and} \quad 0 \quad\mbox{otherwise}
\eeq
and we use also the notation $\{A,B \}_\ast$ for the case of relative + sign.
It is interesting to note that a $Z_2$-grading structure appears in the quantum dimensional weight as well. As discussed in~\cite{AS3}, this seems to suggest that exchanging weights (2 and 0) would be more natural here too.

The $\ast$-bracket formulation \eqref{*5} for the $N=1$ superalgebra therefore yields:
\beq
[\mathcal{\hat{L}}_n,\mathcal{\hat{L}}_m]_\ast=[n-m]\mathcal{\hat{L}}_{n+m}
 +cq^{-1}\, \alpha_{n,m}\mathcal{\hat{J}}_{n+m} \,, \label{superLL*}
\eeq
\beq
\{\mathcal{\hat{G}}_r,\mathcal{\hat{G}}_s  \}_\ast=
(q^{\frac{r-s}{2}}+q^{\frac{s-r}{2}}) q^{1+r+s}\mathcal{\hat{L}}_{r+s}\,, \label{GG*}
\eeq
\beq
[\mathcal{\hat{L}}_n,\mathcal{\hat{G}}_r]_\ast=q^{-n}[\frac{n}{2}-r]\mathcal{\hat{G}}_{r+s}\,. 
\eeq
\beq
 [\mathcal{\hat{L}}_n,\mathcal{\hat{J}}_m]_\ast=-q^{-n}[m] \mathcal{\hat{J}}_{n+m}\,,  \quad
 [\mathcal{\hat{J}}_n,\mathcal{\hat{J}}_m]_\ast=0\,.  \label{JJ*}
\eeq
\beq
[\mathcal{\hat{L}}_n^+,\mathcal{\hat{L}}_m^+]_\ast=[n-m]\mathcal{\hat{L}}_{n+m}^+\,,\quad
[\mathcal{\hat{L}}_n^-,\mathcal{\hat{L}}_m^-]_\ast=[n-m]\mathcal{\hat{L}}_{n+m}^- +c^2\,\alpha_{n,m} \hat{J}_{n+m}\,.   \label{sumLL*}
\eeq

The correspondence with the non-supersymmetric case is as follows: \eqref{superLL*} corresponds to \eqref{*2}, and \eqref{JJ*} corresponds to \eqref{*1} and \eqref{*3}. An anomalous term appears in \eqref{superLL*}. While there is a phase difference in the right-hand side of \eqref{JJ*}, complete agreement can be achieved by redefining $\mathcal{\hat{J}}_n \ra q^n \mathcal{\hat{J}}_n$. To maintain the form of the anomalous term in \eqref{superLL*}, $\alpha_{n,m}$ must also be redefined as $\alpha_{n,m} \ra q^{n+m}\alpha_{n,m}$.

%%%%%%%%%%%%%%%%%%%%%%%%%
%  5.2   N=2 case
%$$$$$$%%%%%%%%%%%%%%%%%%
\subsection{The $N=2$ case}\label{sec:N2*}
\indent

To establish the $N=2$ supersymmetric structure of the super CZ algebra, we decompose the supercharge $G_r$ as follows. This decomposition not only simplifies the algebraic structure but also reveals the underlying supersymmetric properties through the separation of fermionic components.

Let us summarize the final (redefined) expressions. The supercharge representation obtained in MSB form and the superalgebra are as follows:
\beq
\mathcal{\hat{G}}_r=\mathcal{\hat{G}}_r^+ +\mathcal{\hat{G}}_r^-\,,\qquad
\mathcal{\hat{G}}_r^+=\hat{\mu}^{-\nu}\tilde{B}_r\otimes\sigma_1\,,\qquad
\mathcal{\hat{G}}_r^-=c\hat{\mu}^{\nu}\tilde{J}_r\otimes\sigma_2\,,
\label{Gmat}
\eeq
where $\tilde{B}_r$ and $\tilde{J}_r$ are given by \eqref{BJtilde} 
with \eqref{abc_def} and \eqref{e1e2_def}:
\beq
\tilde{B}_r= q^{(1-2\nu)r+\nu+\frac{1}{2}} \hat{B}_{r-\frac{1}{2}}\,,\qquad
\tilde{J}_r=-q^{-r-\frac{1}{2}} \hat{J}_{r+\frac{1}{2}} - q^{r+\frac{1}{2}} \hat{B}_{r+\frac{1}{2}}\,.
\eeq
Under this decomposition, we have the $N=2$ superalgebra:
\beq
\{\mathcal{\hat{G}}_r^\pm,\mathcal{\hat{G}}_s^\pm\}_{(p_0)}=0\,,\label{GGcom}
\eeq
\beq
\{\mathcal{\hat{G}}_r^+,\mathcal{\hat{G}}_s^-\}_{(p_1)}=
q^{2+r+s}q^{(\nu-1)(1+s-r)} \mathcal{\hat{L}}_{r+s}
+q^{(\nu-1)(1+s-r)}q^{\frac{r-s}{2}}[\frac{r-s}{2}]\mathcal{\hat{J}}_{r+s}\,,
\label{hatGG}
\eeq
\beq
[\mathcal{\hat{L}}_n,\mathcal{\hat{G}}_r^\pm]_{(r-\frac{n}{2})}=q^{-n}[\frac{n}{2}-r]\mathcal{\hat{G}}_{r+s}^\pm\,. \label{hatLG}
\eeq
\beq
[\mathcal{\hat{J}}_n,\mathcal{\hat{G}}_r^\pm]_{(r-\frac{n}{2})}=\pm
       q^{n+r+\frac{3}{2}\mp\frac{n+1}{2}}\hat{\mu}\mathcal{\hat{G}}_{n+r}^\pm\,,
\label{hatJG}
\eeq
for any constant $p_0$ and
\beq
p_1=-(r+s)(\nu-1)\,. \label{p1def}
\eeq
The phase factor $p_0$ can be treated as an independent parameter at this stage, though its value will be determined later from the $N=2$ $\ast$-bracket structure. 

Comparing with the quantum superspace formulation, we find that \eqref{hatLG} coincides 
with \eqref{N2LG}, while \eqref{hatGG} corresponds to \eqref{N2GG} under two conditions: 
setting $\nu=1$ and including an additional factor of $q^{\frac{1}{2}}$ on the 
right-hand side. Note that this factor difference can be accommodated through a redefinition of $\mathcal{\hat{G}}_r^\pm$.

Since \eqref{hatGG} reduces to the ordinary anticommutator in the case of $\nu=1$ 
(QSS correspondence), we have to be careful and handle this case separately. 
First, except for \eqref{GGcom} and \eqref{hatGG}, we immediately have the $\ast$-bracket forms:
\beq
[\mathcal{\hat{L}}_n,\mathcal{\hat{G}}_r^\pm]_\ast=q^{-n}[\frac{n}{2}-r]\mathcal{\hat{G}}_{r+s}^\pm\,, \label{N2L*G}
\eeq
\beq
[\mathcal{\hat{J}}_n,\mathcal{\hat{G}}_r^\pm]_\ast=\pm
       q^{n+r+\frac{3}{2}\mp\frac{n+1}{2}}\hat{\mu}\mathcal{\hat{G}}_{n+r}^\pm\,,
\label{N2J*G}
\eeq
where  \eqref{*5} is applied exactly in the same way as the $N=1$ case. 

Apart from the $\nu=1$ case, we have $p_1=\frac{r+s}{2}$ in the case of 
$\nu=\frac{1}{2}$, and we can derive the following $\ast$-bracket expressions for \eqref{GGcom} and \eqref{hatGG}, applying $\eps=+$ and $\eta=-$ to \eqref{X*X} with weight 1:
\beq
\{\mathcal{\hat{G}}_r^\pm,\mathcal{\hat{G}}_s^\pm\}_{\ast}=0\,, \label{N2G*G1}
\eeq
\beq
\{\mathcal{\hat{G}}_r^+,\mathcal{\hat{G}}_s^-\}_\ast=
q^{1+r+\frac{1+s+r}{2}} \mathcal{\hat{L}}_{r+s}
+q^{r-s-\frac{1}{2}}[\frac{r-s}{2}]\mathcal{\hat{J}}_{r+s}\,.  \label{N2G*G2}
\eeq

Here, we should note that these $\ast$-brackets require further refinement.
The $\ast$-brackets for \eqref{N2L*G} and \eqref{N2J*G} do not distinguish 
$\pm$ signs just like the $N=1$ $\ast$-brackets in $\mathscr{M}_S$ \eqref{superM}
, while $\{\mathcal{\hat{G}}_r^\eps, \mathcal{\hat{G}}_s^\eta \}_\ast$ do as seen in \eqref{N2G*G1} and \eqref{N2G*G2}, in other words $\eps\neq\eta$. 
Besides, the phase $x(\eps,\eta)$ of \eqref{X*X} is now $\nu$-dependent.
In these senses, $N=2$ $\ast$-brackets apparently possess 
a certain different structure from the $N=1$ $\ast$-brackets.

In order to make this statement clear, let us introduce the following index $g$,
\beq
g=deg(X_n^{\eps(k)})deg(X_m^{\eta(l)})\,,
\eeq
and define
\beq
\eps_g =\eps^g\,,\quad\, \eta_g=\eta^g\,,
\eeq
\beq
\nu_g= \{2(1-\nu)\}^g\,,
\eeq
where we have defined $\nu_g=1$ for $g=0$. 
The phase factor $x(\eps,\eta)$ is then replaced by
\beq
x_g(\eps,\eta)=\nu_g \frac{nl\eta_g - mk\eps_g}{2}\,. \label{xg}
\eeq
This phase definiton instead of \eqref{X*X} covers all of Eqs.\eqref{N2L*G}-\eqref{N2G*G2} as well as the $\nu=1$ case. 
In order to reproduce the $N=1$ case, we have to force $g=0$ for all $\ast$-brackets, 
yielding $\eps_g=\eta_g=\nu_g=1$. However, \eqref{GG*} leads to $g=1$, 
we therefore recognize that the $N=2$ $\ast$-brackets are different from the $N=1$ ones.

In order to establish a complete $N=2$ $\ast$-bracket formalism that is consistent with $N=1$, we have to take into account the dual components $\mathcal{\hat{L}}_n^{(\pm)}$, where 
$\mathcal{\hat{L}}_n^{(+)}=\mathcal{\hat{L}}_n$ and $\mathcal{\hat{L}}_n^{(-)}$ defined by 
the interchange $q\ra q^{-1}$ from $\mathcal{\hat{L}}_n^{(+)}$.
This structure parallels that of the $CZ^\ast$ algebra, suggesting that \eqref{X*X} 
in $\mathscr{M}_S$ should be extended to $\mathscr{M}_{S_2}$:
\beq
X_n^{(k)} \in \mathscr{M}_{S_2}\{\mathcal{\hat{L}}_n^\pm, \mathcal{\hat{J}}_n, \mathcal{\hat{G}}_r^\pm \} \,.
\eeq

%%%%%%%%%%%%%%%%%%%%%%%%%%%%%%%%%%%%%%%%%%%%%%%
%  6.    super CZ in TBM-spin system
%%%%%%%%%%%%%%%%%%%%%%%%%%%%%%%%%%%%%%%%%%%%%%%
\setcounter{equation}{0}
\section{Super CZ Algebra in TBM-Spin Systems}\label{sec:TBM}
\indent

In this section, we demonstrate the realization of super CZ algebras in a TBM discrete system with spin interactions, building upon the continuous system results from the previous sections. Since the TBM Hamiltonian can be expressed in terms of cyclic matrix representations of $CZ^\ast$~\cite{AS2,AS2e}, we can derive the cyclic matrix representation of super $\,\scz$ by utilizing the mapping relationship between MT operators and cyclic matrices, starting from the MSB super $\,\scz$ operators.

In discrete lattice systems such as tight-binding models~\cite{WZ}-\cite{MB}, it is appropriate to consider an alternative representation: the cyclic matrix representation~\cite{NQ} with parameters $a_\pm$ and $b$,
\beq
{\mathrm L}^\pm_n=\mp\left( \frac{1-{\mathrm Q}^{\pm2}}{q-q^{-1}}
+A_n^{\pm} {\mathrm Q}^{\pm2} \right)
{\mathrm H}^n \,,\qquad A_n^{\pm}=a_\pm+b(q^{\pm2n}-1)\,,    \label{QHCZ}
\eeq
expressed in terms of Weyl basis matrices ${\mathrm H}$ and ${\mathrm Q}$ satisfying the commutation relation
\beq
 {\mathrm H} {\mathrm Q}= q{\mathrm Q} {\mathrm H}\,,
\eeq
where the matrix elements are defined as
\beq
H_{jk}=\delta_{k+1, j}\,,\quad Q_{jk}=q^{j-1}\delta_{jk}\,,\quad\mbox{for}\quad j,k \in [1,N] \quad(\mbox{mod}\,N) \,. \label{HQdef}
\eeq
Selecting $a_\pm$ and $b$ as follows:
\beq
a_\pm=0\,,\quad b=-1/(\qqi)\,,
\eeq
and applying the correspondence to \eqref{QHCZ}
\beq
z \lra {\mathrm H}\,,\quad q^{\mp2z\partial} \lra {\mathrm Q}^{\pm2} \,,
\label{MT2HQ}
\eeq
we can verify that this representation coincides with the $q$-difference representation \eqref{Ln_diff}. The matrix form of the scaling operator can be expressed as:
\beq
S_0^\pm=1\pm(\qqi){\mathrm L}_0^\pm = \{1-A_0^\pm (\qqi)\} {\mathrm Q}^{\pm2} \,.
\label{matS0}
\eeq

We begin by verifying the mapping relationship \eqref{MT2HQ} between differential MT operators and cyclic matrices. The general form of matrix operators for the $CZ^\ast$ algebra in the TBM system \eqref{QHCZ} can be written as:
\beq
{\mathrm L}^\pm_n={\mathrm B}_n^\pm + g_n^{\pm} {\mathrm J}_n^\pm\,,\quad
g_n^{\pm}=\pm q^{\mp2n}\frac{1-A_n^{\pm}c}{\qqi}\,,\label{LisBJ}
\eeq
where $g_n^{\pm}$ is the coupling coefficient $g_n^{CZ}$ defined in \eqref{CZgen}, and for $CZ^-$ is obtained by the transformation $q\ra q^{-1}$. The matrices ${\mathrm B}_n^\pm$ and ${\mathrm J}_n^\pm$ are expressed using \eqref{HQdef}:
\beq
{\mathrm B}_n^\pm=\frac{\mp1}{\qqi}{\mathrm H}^n\,,\quad
{\mathrm J}_n^\pm=\frac{1}{\qqi}{\mathrm H}^n{\mathrm Q}^{\pm2}
=\mp{\mathrm B}_n^\pm {\mathrm Q}^{\pm2}\,. \label{BJmat}
\eeq
Note that while the coefficients differ slightly from Eq.(4.15) in \cite{AS2,AS2e}, this is due to normalization adjustments made to maintain the algebraic relations satisfied by $T_n^{(0)}$ and $T_n^{(2)}$.

The matrix representation of the scaling operator from \eqref{matS0} is (setting $a_\pm=0$ for simplicity\footnote{This is merely a convenient choice to make $\hat{\mu}$ and ${\mathrm S}_0^\pm$ correspond with coefficient 1. Since we must set $b=0$ to satisfy $CZ^\ast$~\cite{AS2,AS2e}, setting $a\pm=0$ simultaneously would make $A_n^\pm=0$. Therefore, we should actually have $a_\pm\not=0$.}):
\beq
{\mathrm S}_0^\pm=1\pm(\qqi){\mathrm L}_0^\pm={\mathrm Q}^{\pm2}\,.
\label{matS00}
\eeq
For the MT representation of $CZ^\pm$, from \eqref{That} and \eqref{Lpm1}, we obtain:
\beq
\hat{\mu}^{\pm1}=1\pm(\qqi)\hat{L}_0^\pm =cq^{\pm2\Delta}\Op{T}{0}{\pm2}
=q^{\mp2z\partial}\,.  \label{mudiff}
\eeq
This reveals that the differential operator $\hat{\mu}$ corresponds to the matrix ${\mathrm S}_0^+$. Furthermore, the $CZ$ differential operators $\hat{B}_n$ and $\hat{J}_n$ defined in \eqref{LbyBJ} etc. extended to $CZ^\pm$ are:
\begin{align}
&\hat{L}_n^\pm=\hat{B}_n^\pm\pm \hat{J}_n^\pm  \label{LbyBJ2}\\
&\hat{B}_n^\pm=\mp\Op{T}{n}{0}=\frac{\mp z^n}{\qqi} \\
&\hat{J}_n^\pm=q^{n+2\Delta}\Op{T}{n}{\pm2}=z^n\frac{q^{\mp2z\partial}}{\qqi}
=\mp\hat{B}_n^\pm \hat{\mu}^{\pm1}\,.
\end{align}
One can see that these operators are related to \eqref{BJmat} and \eqref{matS00} 
by the correspondence \eqref{MT2HQ}, and satisfy the $\ast$-commutation relations using the $\ast$-commutator \eqref{X*X} of $\mathscr{M}_T$:
\begin{align}
&[{\mathrm B}_n^\pm,{\mathrm B}_m^\pm]_\ast=[n-m]{\mathrm B}_{n+m}^\pm\,,\\
& [{\mathrm B}_n^\eps,{\mathrm J}_m^\eta]_\ast=-q^{-n\eps}[m] {\mathrm J}_{n+m}^\eta\,, \quad
 [{\mathrm J}_n^\pm,{\mathrm J}_m^\pm]_\ast=0\,.  \label{hatBJJJ2}
\end{align}
This demonstrates that the MT representation $(\hat{B}_n,\hat{J}_n,\hat{\mu}^{\pm1})$ and cyclic matrix representation $({\mathrm B}_n,{\mathrm J}_n,{\mathrm S}_0^\pm)$ can be transformed into each other via substitution \eqref{MT2HQ}.

To determine the concrete form of the supercharge \eqref{Gform}, we must specify a value for the undetermined parameter $\nu$. Although examining $\nu=\frac{1}{2}$ would reveal furthermore details of $N=2$ $\ast$-product structure or a connection with $GL_q(1,1)$, we select $\nu=1$ here to achieve simpler coefficient forms in \eqref{abc_def}. By setting $\nu=1$ in the $\nu$-dependent parameters and operators (derived from \eqref{BJtilde}, \eqref{abc_def}, \eqref{p1def}, \eqref{matS00} and \eqref{mudiff}), we obtain:
\[
\gamma=\frac{3}{2}-r\,,\quad p_1=0\,,\quad
\hat{\mu}^{\pm\nu}\ra{\mathrm Q}^{\pm2}\,,\quad 
\tilde{B}_r\ra \tilde{{\mathrm B}}_r=
\frac{-q^{\frac{3}{2}-r}}{\qqi}{\mathrm H}^{r-\frac{1}{2}}\,.
\]
The remaining parts can be obtained through direct substitution $(\hat{B}_n,\hat{J}_n)\ra({\mathrm B}_n,{\mathrm J}_n)$ and application of the matrix representation \eqref{BJmat}. 
As a result of this substitution, 
$(\mathcal{\hat{L}}_n,\mathcal{\hat{G}}_r^\pm,\mathcal{\hat{J}}_n)$ 
is redefined in terms of 
$(\mathbb{L}_n,\mathbb{G}_r^\pm,\mathbb{J}_n)$, 
yielding the matrix representation for each generator: 
\begin{align}
&\mathbb{G}_r^+=\frac{-q^{r+\frac{1}{2}}}{\qqi}{\mathrm H}^{r-\frac{1}{2}}
{\mathrm Q}^{-2}\otimes \sigma_1\,, \label{bbG+}\\
&\mathbb{G}_r^-=-q^{-r-\frac{1}{2}} {\mathrm H}^{r+\frac{1}{2}}
{\mathrm Q}^{2}(q^{-2r-1}{\mathrm Q}^2-1 ) \otimes \sigma_2\,,\label{bbG-}
\end{align}
\beq
\mathbb{L}_n=q^{-2n}\mathbb{L}_n^+ \otimes\sigma_2\sigma_1
+\mathbb{L}_n^- \otimes\sigma_1\sigma_2\,, \label{bbL}
\eeq
\beq
\mathbb{L}_n^+=\frac{{\mathrm H}^n }{\qqi}(-1+q^{-2n}{\mathrm Q}^2)\,,\quad
\mathbb{L}_n^-=\frac{{\mathrm H}^n }{\qqi}(-1+q^{-1-n}{\mathrm Q}^2)\,,
\label{bbLpm}
\eeq
\beq
\mathbb{J}_n= q{\mathrm H}^n{\mathrm Q}^2 \otimes\sigma_1\sigma_2\,.
\label{bbJ}
\eeq
These expressions satisfy the $N=1$ and 2 super CZ algebra relations \eqref{superLL}-\eqref{N2G*G2}. As a specific example, \eqref{hatGG} simplifies to ordinary anticommutators, establishing correspondence with the QSS relation \eqref{N2GG} (noting that $\mathbb{J}_n$ carries a scale factor of $q$, and, 
as commented below \eqref{hatLG}, also the factor of $q^{\frac{1}{2}}$):
\beq
\{\mathbb{G}_r^\pm,\mathbb{G}_s^\pm\}=0\,,\quad
\{\mathbb{G}_r^+,\mathbb{G}_s^-\}=q^{2+r+s} \mathbb{L}_{r+s}
+q^{\frac{r-s}{2}}[\frac{r-s}{2}]\mathbb{J}_{r+s}\,, \label{nu1GG}
\eeq
where $p_0=0$ has been chosen in \eqref{GGcom}. 

The case of $\nu=\frac{1}{2}$ is presented in Appendix~\ref{app:Weyl}, where 
we notice slight changes in \eqref{bbG+}-\eqref{bbL}. Since $p_1\neq0$ in the case, 
we have different relations from \eqref{nu1GG}: 
\beq
\{\mathbb{G}_r^\pm,\mathbb{G}_s^\pm\}_{(p_0)}=0\,,\quad
\{\mathbb{G}_r^+,\mathbb{G}_s^-\}_{(\frac{r+s}{2})}=q^{\frac{1}{2}(3+3r+s)} \mathbb{L}_{r+s}
+q^{\frac{r-s-1}{2}}[\frac{r-s}{2}]\mathbb{J}_{r+s}\,.
\eeq
One may choose $p_0$ as $p_0=p_1=\frac{r+s}{2}$.

%\vskip\baselineskip
%%%%%%%%%%%%%%%%%%%%%%%%%%%%%%%%%%%%%%%%%%%%%%%%%%%%%%%%%%%%%%%%%%%%%%
%                                     7.  Conclusions 
%%%%%%%%%%%%%%%%%%%%%%%%%%%%%%%%%%%%%%%%%%%%%%%%%%%%%%%%%%%%%%%%%%%%%%
\setcounter{equation}{0}
\section{Conclusions and Outlook}\label{sec:end}
\indent

In this paper, we have developed both the MT representation and cyclic matrix representation of super CZ algebras in Bloch electron systems with Zeeman effects. Through the introduction of magnetic fields, noncommutative structures naturally emerge as quantum plane pictures, while spin interactions generate super CZ algebras analogous to those constructed on QSS. 
The SSM correspondence between QSS and MSB demanded careful analysis of the duality inherent in QSS scaling operators. This intrinsic duality necessitated the introduction of mixing states that bridge bosonic and Grassmann representations, ultimately leading to the realization of Type 3 super CZ algebra through a combination of MT operators and spin matrix bases. Most significantly, we have established a comprehensive $\ast$-bracket formalism that unifies these structures and illuminates their fundamental properties.

The super CZ algebra closes by itself without embedding into other algebras such as 
the Virasoro algebra. This simply reflects the intrinsic symmetry structure of the 
Bloch electron system formulated within the quantum superspace framework, and does 
not imply the emergence of additional physical degrees of freedom.

We begin by reviewing our problem formulation. The foundational elements for addressing this problem were organized in Section \ref{sec:QSS} and Apendix \ref{sec:SV}. In Appendix \ref{sec:SV}, we reviewed the correspondence between spin Grassmann bases in electron spin systems under static magnetic fields and Grassmann coordinates in superspace. Through this correspondence, we demonstrated how Virasoro super generators and supercharge could be systematically constructed as block matrix representations with spin Grassmann bases. This framework represents a conventional quantum system without MT operators or quantum space concepts, where supersymmetric structure emerges solely through weak magnetic fields. To establish noncommutative structure while preserving supersymmetry, we required either the QSS approach or the introduction of MT operators in strong magnetic fields. The central challenge lay in bridging these two perspectives.

The QSS approach provides rigorous mathematical definitions of possible super CZ algebras, yet its connection to physical systems remains elusive. In contrast, while the MT approach offers clear physical interpretations, it lacks a systematic framework for combining MT operators to construct super CZ algebras. We have addressed this dichotomy by developing a unified framework that consistently incorporates both approaches.

In Section \ref{sec:SCZ}, we examined three established types of super CZ algebras constructed on QSS, with our analysis revealing fundamental insights into their characteristics. Type 1 represents the most direct formulation, where the CZ algebra's right-hand side is free of $U(1)$ current terms. However, since its supercharge algebra fails to manifest as a pure super CZ generator (due to additional $U(1)$ current terms), we determined it unsuitable for our purposes and excluded it from consideration. The supercharge realization exhibits two additional variations (Types 2 and 3). However, Type 2, analogous to type 1, incorporates extraneous $U(1)$ terms (as evidenced in \eqref{GrGs1} and \eqref{GGtab2}). We finally decide that our desired formulation is Type 3, where the supercharge algebra's right-hand side comprises solely super CZ generators. 

To implement this structure, the Virasoro component corresponding to the CZ algebra requires modification, as demonstrated in \eqref{CZtwist}. While the excess $U(1)$ terms disappear from the supercharge algebra, they are systematically incorporated within the CZ algebra. Beyond this structural modification, the formulation maintains its equivalence to Type 1, employing identical supercharge forms but with a modified scaling operator. The resulting anomaly in the super algebra exhibits proportionality to $c=\qqi$ and vanishes at $q=1$. To differentiate this modified structure from the conventional CZ algebra, we designate it as $\,\scz$.

A particularly significant feature of Type 3 super $\,\scz$ is its accommodation of $N=2$ decomposition and its natural realization in electron spin systems. This characteristic opens intriguing avenues for future investigation into other QSS beyond $GL_q(1,1)$, or other Types.

%  電子スピン系では CZ と scz がblock行列要素に現れる。
%   $\pm$で表示したが、$CZ^\pm$ と混同しないように注意。
%  \hat{\mu} はMTのスケール演算子    \mu はQSのスケール演算
%
In Section \ref{sec:MSB}, we explored the extension of QS-MT correspondence to supersymmetric case and developed the construction of super CZ (in MT) from super CZ (in QSS) through operator mixing. The central challenge lay in establishing a precise mapping from QSS to SM space's Grassmann basis. The complexity was further compounded by the subtle nature of supersymmetric operator manifestations through combinations of $\hat{B}_n$ and $\hat{J}_n$.

Our solution proceeded in steps. First, in pure bosonic case, we confirmed the correspondence between QS differential operators $(x,\partial_x,\mu)$ and MT $q$-differential operators $(z,\partial_q,\hat\mu)$, establishing that $CZ_{QS}$ generators could be recast as $CZ_{MT}$ generators ($\hat{B}_n$,$\hat{J}_n$). Next, while maintaining this relationship, we needed to map super $CZ_{QSS}$ to SM space's MT representation. This required careful consideration of how QSS scaling operator $\mu$'s MT counterpart $\hat\mu$ manifests in SM basis. While pure bosonic QS's $\mu$ only needed bosonic representation, QSS's $\mu$ having two equivalent representations - bosonic and Grassmann - made the problem non-trivial. 
Note that this characteristic of $\mu$ applies to Type 1 and Type 2 cases, 
but it similarly holds true for $\lambda$ in Type 3 up to the distortion factor $q^{-1}$ for 
down-spin component as seen in \eqref{lambda_mat}. 
After all, $\lambda$ maps to $\hat{\mu}$ as understood from \eqref{scale2}.

Given that the generators' structures need not maintain strict equivalence, we introduced a framework permitting linear transformations. Through the introduction of mixed state $\mathfrak{J}_n$ (although its underlying emergence mechanism remains to be elucidated), we established the MT correspondence for Type 3. The necessity for mixing emerges naturally from the dual nature of QSS, as analyzed in Section \ref{sec:mix}. This duality manifests in two distinct aspects. Primarily, the duality of $\mu$: its bosonic representation maps to $\hat{\mu}$, while its Grassmann representation corresponds to the unit matrix in MSB representation. Secondarily, reflecting this fundamental duality, the twisted operator $\mathfrak{J}_n$ derived from $J_n$ exhibits analogous dual characteristics. These observations provide no compelling reason to exclude mixing possibilities in the correspondence between QSS $(B_n,\mathfrak{J}_n)$ and MSB space $(\hat{B}_n,\hat{J}_n)$. The same argument applies to the discrete version 
$(\mathrm{B}_n,\mathrm{J}_n)$ investigated in Section~\ref{sec:TBM}.

The inherent ambiguity extends further: just as the origin of unit matrix representations in QSS (whether from 1 or $\mu$) remains indeterminate, the precise correspondence of MSB space operator $\hat{B}_n$ to either $B_n$ or $\mathfrak{J}_n$ in QSS cannot be definitively established. This fundamental ambiguity might indicate the existence of an as-yet-unidentified physical mechanism underlying the twisting phenomenon.

Type 3 super CZ encompasses both CZ and $\,\scz$ sectors, featuring an anomaly term proportional to $c$ that vanishes at $q=1$. This inherent dual sector structure potentially accounts for the insufficiency of simple substitution correspondence and the necessity of mixing. The question of whether this phenomenon is unique to $GL_q(1,1)$ or extends to general QSS remains unresolved. Although we excluded Types 1 and 2 from our analysis, analogous mixing approaches might illuminate their MSB representations.

The mixing matrix elements are related to the coefficients $g_n^{CZ}$ and $g_n$, 
and their freedom of choice might facilitate the realization of additional superalgebra types, as demonstrated in \cite{super3,HHT}. Furthermore, investigation of connections with other MT-based super Virasoro algebra deformations \cite{JS,KS} could present a promising direction for future research.

Analysis in Section \ref{sec:*form} identified two fundamental issues concerning super $\ast$-brackets:
\begin{enumerate}
\item The consistency requirements for $N=2$ $\ast$-brackets (which may necessitate $CZ^\ast$ for a coherent formulation with $N=1$ $\ast$-brackets) \par\indent
\item The potential alignment between $Z_2$-grading degree (=0) and weight (=$\pm2$) for enhanced mathematical elegance
\end{enumerate}
These issues require resolution in conjunction with the weight-related questions noted in \cite{AS3}:
\begin{enumerate}\setcounter{enumi}{2}
\item The apparent discrepancy between weight $k=\pm2$ of $CZ^\pm$ generators $L_n^\pm$ and weight $\pm k$ of scaled algebra $\mathcal{CZ}$ generators $L_n^{\pm(k)}$. Specifically, although $L_n^{\pm(0)}=L_n^\pm$ suggests $k=0$, $CZ^\pm$ weights necessarily take values $\pm2$
\item The phase sign inversion observed in $\ast$-brackets between $CZ$ and scaled $\mathcal{CZ}$ algebras
\end{enumerate}
A possible resolution of these four issues might be achieved through weight exchange involving role reversal between $T_n^{(0)}$ and $T_n^{(\pm2)}$.

The theoretical insights developed throughout this investigation have begun to unify previously fragmentary understanding of these mathematical structures. Our analysis demonstrates increasing coherence in the relationships among seemingly disparate superalgebras, their quantum space representations, and their physical manifestations. 
The systematic characterization of these algebraic properties establishes a theoretical framework that not only illuminates existing problems but also suggests new directions for investigation.

The mathematical structures established in this work --- specifically the $\ast$-bracket formalism incorporating $Z_2$-grading, the mixing mechanism connecting QSS and MT representations, and the fundamental role of quantum dimensional weights --- provide a rigorous foundation for exploring quantum geometry and its physical realizations such as
a discrete Hamiltonian system investigated in Section~\ref{sec:TBM}. Although significant questions persist, particularly concerning the extension of our results beyond $GL_q(1,1)$ and the fundamental nature of the mixing mechanism, this work establishes systematic approaches for investigating the intricate relationships between supersymmetry, quantum deformation, and physical systems.

The framework developed here might also provide insights for other areas of physics. For instance, within the context of nonextensive statistical mechanics~\cite{TS}, the study of superstatistics with q-deformed structures~\cite{WSC} has led to various applications in physical systems, including microcanonical ensemble formulations, quantum Hall effects, and deformed quantum mechanical systems~\cite{WSC2}-\cite{WSC4}. These developments suggest potential connections between our mathematical framework and broader physical applications, particularly in systems exhibiting quantum deformation characteristics.

%%%%%%%%%%%%%%%%%%%%%%%%%%%%%%%%%
%  CRediT statement
%%%%%%%%%%%%%%%%%%%%%%%%%%%%%%%%%
\section*{Declaration of generative AI and AI-assisted technologies in the writing process}
During the preparation of this work the author used Claude 3.5 Sonnet in order to improve grammar and enhance language expression. 
After using the service, the author reviewed and edited the content as needed and takes full responsibility for the content of the publication.

\section*{CRediT authorship contribution statement} 
 \textbf{Haru-Tada Sato:} Writing-Original Draft, Conceptualization, Methodology, Investigation, Validation.

\section*{Declaration of competing interest}
The author declares that we have no known competing financial interests or personal relationships that could have appeared to influence the work reported in this paper.

%\section*{Acknowledgements}
%The author would like to thank Professor N. Aizawa for the valuable discussions and %comments on this paper.

\section*{Data availability}
No data was used for the research described in the article. 

%%%%%%%%%%%%%%%%%%%%%%%%%%%%%%%%%%%%%%%%%%%%%%%%%%%%%%%%%%%%%%%%%%%
%                               Appendix A
\appendix
%%%%%%%%%%%%%%%%%%%%%%%%%%%%%%%%%%%%%%%%%%%%%%%%%%%%%%%%%%%%%%%%%%%
\setcounter{equation}{0}
%%%%%%%%%%%%%%%%%%%%%%%%%%%%%%%%%%%%%
%  A     Definitions
%%%%%%%%%%%%%%%%%%%%%%%%%%%%%%%%%%%%%
\section{Notation and Conventions}\label{sec:def}
\indent

In this appendix, we summarize the notations and abbreviations frequently used throughout the paper. This list is intended as a quick reference for readers. 
Precise QSS operator definitions for super CZ algebras are summarized in 
Appendix~\ref{sec:3types}, while here we focus on the operators relevant to Type 3, 
which plays the central role in the present work.

\subsection{Abbreviations}

\begin{description}
\item[QS]: Quantum space (bosonic sector).
\item[QSS]: Quantum superspace with noncommutative coordinates $(x,\theta)$.
\item[SS]: Standard superspace.
\item[MT]: Magnetic translation operators.
\item[MSB]: Magnetic spin-matrix basis (see \eqref{MSbase}).
\item[SSM]: Superspace and spin matrix correspondence (see \eqref{grass}).
\item[CZ]: Curtright–Zachos algebra.
\item[Super CZ algebra]: Supersymmetric CZ algebra. Three types exist. See Appendix~\ref{sec:3types}.
\item[FFZ]: Fairlie-Fletcher-Zachos algebra.
\item[TBM]: Tight binding model.
\end{description}

\subsection{Notation}
We summarize the notation used for various representations of the super CZ generators.
\par\noindent
\subsubsection*{CZ generator}
\begin{description}
\item[$L_n,B_n,J_n$]: QS representation (denoted as $CZ_{QS}$)
\item[$\hat{L}_n,\hat{B}_n,\hat{J}_n$]: MT (differential operator) representation (denoted as $CZ_{MT}$)
\item[$\mathrm{L}_n,\mathrm{B}_n,\mathrm{J}_n$]: Weyl matrix representation
\end{description}
\subsubsection*{Supersymmetric case}
\begin{description}
\item[$L_n,G^\pm_n,B_n,J_n$]: QSS representation (Bosonic part is denoted as $CZ_{QSS}$)
\item[$\mathcal{\hat{L}}_n,\mathcal{\hat{G}}^\pm_n,\mathcal{\hat{J}}_n$]: MSB representation. 
$\mathcal{\hat{L}}^\pm_n$ define upper/lower diagonal parts of $\mathcal{\hat{L}}_n$.
\item[$\mathbb{L}_n,\mathbb{G}^\pm_n,\mathbb{J}_n$]: supersymmetric Weyl matrix 
 representation. $\mathbb{L}^\pm_n$ define upper/lower diagonal parts 
of $\mathbb{L}_n$.
\end{description}

In order to avoid confusion due to the proliferation of notation, the bosonic part of the CZ algebra in the supersymmetric CZ formulation is taken to be $CZ^+$ by default. 
Note that the notation $L^\pm_n$ (for $CZ^\pm$ generators) does not necessarily 
imply the upper/lower diagonal parts in supersymmetric matrix representations, 
$\mathcal{\hat{L}}^\pm_n$ or $\mathbb{L}^\pm_n$.

\subsection{CZ algebras}
\begin{description}
\item[$CZ^\pm$]: 
Dual pair under the transformation $q\ra q^{-1}$; given by \eqref{*2}~\cite{AS2,AS2e}. 
Unless otherwise specified, CZ refers to $CZ^+$.
Explicitly writing down \eqref{*2}, using \eqref{CZCZ}-\eqref{X*X}, we have
\beq
CZ^+:\qquad
[L^+_n,L^+_m]_{(m-n)}=[n-m]L^+_{n+m}\,, \label{app+}
\eeq
\beq
CZ^-:\qquad
[L^-_n,L^-_m]_{(n-m)}=[n-m]L^-_{n+m}\,. \label{app-}
\eeq

\item[$CZ^\ast$]: 
Extended CZ algebra constructed from both of $L^\pm_n$ \cite{AS2,AS2e}. From \eqref{CZCZ}, we need the following relation in addition to \eqref{app+} and \eqref{app-} for defining $CZ^\ast$: 
\beq
[L^+_n,L^-_m]_{(n+m)}=q^{-m}[n]L^+_{n+m}-q^n[m]L^-_{n+m}\,.
\eeq

\item[$\mathcal{CZ}$]: 
Another extended CZ algebra. There are versions such as $\mathcal{CZ}^\pm$ and $\mathcal{CZ}^\ast$, of which generators are denoted as $L^{\pm(k)}_n$ in \eqref{Lnpmk}. 
$CZ^\ast$ is composed of $k=0$ generators of $\mathcal{CZ}$, namely given by 
$L^{\pm(0)}_n = L^\pm_n$. 
According to a rescaling formula (Eq.(3.29) in \cite{AS3}), 
we introduce instead of \eqref{Lnpmk}: 
\beq
\tilde{L}^{\eps(k)}_n:= q^{-2\eps\Delta}L^{\eps(k)}_n
=-\eps q^{-2\eps\Delta}\hat{T}_n^{(-\eps k)}+\eps q^{\eps n}
\hat{T}_n^{(-\eps k+2\eps)}\,,
\eeq
where $\eps$ stands for the $\pm$ sign. This generator satisfies the following 
$\ast$-bracket algebra, very similar to \eqref{CZCZ}:
\beq
[\tilde{L}^{\eps(k)}_n,\tilde{L}^{\eta(l)}_m]_\ast=
q^{\eta m}[n]\tilde{L}^{\eps(k+\eps\eta l-2\eps\eta)}_{n+m}
-q^{\eps n}[m]\tilde{L}^{\eta(l+\eps\eta k-2\eps\eta)}_{n+m}\,,\label{newCZ}
\eeq
where the sign of $x(\eps,\eta)$ in \eqref{X*X} should be inverted. 
Note that \eqref{newCZ} reduces $CZ^\ast$ for not $\tilde{L}^{\pm(0)}_n$ 
but $\tilde{L}^{\pm(2)}_n$ because of the inversion of $\ast$-product phase 
$x(\eps,\eta)$. For further details, refer to Section 3 in \cite{AS3}. 

\item[$\,\scz$]: 
CZ algebra \eqref{CZtwist} with an anomalous term (taken with the $CZ^+$ basis in this paper). 
\beq
[L'_n,L'_m]_{(m-n)}=[n-m]L'_{n+m}+a_{n,m}J_{n+m}\,, \qquad
\eeq
where $a_{n,m}$ given by \eqref{a_nm}. There are additional generators $J_n$, which satisfy \eqref{FFU1q} and \eqref{type1LL}
\beq
[J_n,J_m]_{(m-n)}=0\,,\qquad
[L'_n,J_m]_{(m-n)}=-q^{-n}[m]J_{n+m}\,.
\eeq
\end{description}

%%%%%%%%%%%%%%%%%%%%%%%%%%%%%%%%%%%%%%%%%%%%%%%%%%%%%%%%%%%%%%%%%%%
%                               Appendix  B 
%%%%%%%%%%%%%%%%%%%%%%%%%%%%%%%%%%%%%%%%%%%%%%%%%%%%%%%%%%%%%%%%%%%
\setcounter{equation}{0}
%%%%%%%%%%%%%%%%%%%%%%%%%%%%%%%%%%%%%
%  B     Super CZ (Type 1-3)
%%%%%%%%%%%%%%%%%%%%%%%%%%%%%%%%%%%%%
\section{Summary of Super $CZ$ Algebras (Types 1–3)}\label{sec:3types}
\indent

This appendix, based on \cite{superCZ,superCZ2}, systematically organizes the three super CZ algebras.
Those papers refer to formulations based on quantum $OSP(1,2)$ and $GL(1,1)$, but the bicovariant calculus determined by these quantum groups yields, at least in the case of the $1+1$ dimensional superspace, the same commutation relations. Since these works focus on constructing algebras via differential operators rather than emphasizing the underlying background structures, either choice is acceptable.

We make use of the following $GL_q(1,1)$ covariant 
quantum superspace~\cite{superCZ,superCZ2}:
\begin{align}
&(\theta)^2=(\partial_\theta)^2=0, \quad x \theta =  q\theta x,
\quad \delx \delt =  q^{-1}\delt \delx,        \nn   \\
&\delx x=1+q^{-2} x \delx, \quad \delt \theta 
           = 1- \theta \delt +(q^{-2}-1)x \delx,  \\
&\delx\theta=q^{-1}\theta\delx, \quad \delt x=q^{-1}x\delt, \nn
\end{align}
where $x$ and $\theta$ are the bosonic and fermionic (noncommutative) 
coordinates. The scaling operator $\lambda$ is realized by 
\beq
   \lambda = \mu+(q^{2a}-1)\theta\delt \ , \label{lam}
\eeq
with another scaling operator 
\beq
\mu=\delx x-x\delx=\delt\theta+\theta\delt=1+(q^{-2}-1)x\delx\,.
\eeq
This leads to the following scaling relations:
\begin{align}
&\lambda x=q^{-2}x \lambda, \qquad 
  \lambda \delx =q^2 \delx\lambda,  \qquad
\lambda\theta=q^{2a}\theta\lambda,\qquad  
\lambda\delt=q^{-2a}\delt\lambda, \\
&\mu x=q^{-2}x \mu, \qquad \mu \delx =q^2 \delx \mu, \qquad
[\mu,\theta]=[\mu,\delt]=0.
\end{align}

Since these intricate commutation relations complicate various calculations, it is useful to keep the following formulas in mind ($n\in \mathbb{Z}$ or $\mathbb{Z}+\tfrac{1}{2}$) 
\begin{align}
 & \delx x^n = q^{-2n} x^n \delx + q^{-n+1} [n] x^{n-1}, \\
 & \theta \delt x^n = q^{-2n} x^n \theta \delt, \qquad \theta \delt \delx = q^{2} \delx \theta \delt, 
\end{align}
\begin{align}
&\delx x^n  =  - q^{-2n+1}x^{n-1}{\mu-q^{2n}\over q-q^{-1}}, 
\qquad n\not=0,\\
&(\theta\delt)^n  =  \mu^{n-1}\theta\delt. 
\end{align} 

%%%%%%%%%%%%%
%    Type 1
%%%%%%%%%%%%%
\subsection{Type 1 super CZ algebra}

Type 1 and 2 were originally studied in \cite{superCZ2}. 
The general form of Virasoro counterpart $L_n$ is defined by \eqref{CZgen} with 
the notation \eqref{Bn}, and 
we choose the coupling $g_n$ as \eqref{gn_1} for Type 1 super CZ algebra:
\beq
 L_n=B_n-g_n J_n\,,\qquad g_n=\frac{1}{[2]}q^{-n}[n+1] \,,
\eeq
where
\beq
B_n=-q^{-1}x^{n+1}\partial_x\,,\qquad J_n=x^n\theta\partial_\theta\,. 
\eeq

The supercharge $G_r$ is defined by \eqref{Gtype1}:
\beq
G_r=\mu^{-\frac{1}{2}}x^{r+\frac{1}{2}}(\partial_\theta-\theta\partial_x)\,, 
\eeq
and $\mu$ is related to the zero mode $L_0$ by \eqref{mubyL0}:
\beq
\mu=1+(q-q^{-1})L_0\,.
\eeq

Algebraic relations are summarized in Table~\ref{tab:compare}. Type 1 superalgebra 
consists of \eqref{type1LL}, \eqref{type1GG} and \eqref{LnGr1}:
\beq
[L_n,L_m]_{(m-n)}=[n-m]L_{n+m}\,,\qquad[L_n,J_m]_{(m-n)}=-q^{-n}[m]J_{n+m}\,,
\eeq
\beq
\{G_r,G_s\}_{(s-r)}=q^{r+s+2}(q^{s-r}+q^{r-s})B_{r+s}
-q^{\frac{3}{2}}(q^s[s+\frac{1}{2}]+q^r[r+\frac{1}{2}])J_{r+s}\,, 
\eeq
\beq
[L_n,G_r]_{(r+\frac{1}{2}-n)}=q^{-n}[n-r-\frac{1}{2}]G_{n+r} 
+q^{n-r-\frac{1}{2}-2n(n+r+\frac{3}{2})}\frac{[1-n]}{[2]}\,\mu^{-n}G_{n+r}\mu^{n+1}\,.
\eeq

%%%%%%%%%%%%
%   Type 2
%%%%%%%%%%%%
\subsection{Type 2 super CZ algebra}

The general form of Virasoro counterpart $L'_n$ is defined by \eqref{CZTgen} with 
the notation \eqref{Bn}, and 
we choose the coupling $g'_n$ as \eqref{Gtype2} for Type 2 super CZ algebra:
\beq
 L'_n=B_n-g'_n J_n\,,\qquad g'_n=q^{-\frac{n}{2}}[\frac{n}{2}] \,.
\eeq
In this case as well as in Type 3, we have introduced the dashed notation such as $L'_n$, 
since the bosonic CZ algebra gives rise to anomalous terms. We thus denote the algebra 
as $\,\scz$. 
The supercharge $G_r$ is defined by \eqref{Gtype2}
\beq
G_r=\mu^{-\frac{1}{2}}x^r(\partial_\theta-x\theta\partial_x)\,,
\eeq
and $\mu$ is related to the zero mode $L'_0$ instead of \eqref{mubyL0}:
\beq
\mu=\delx x-x\delx=1+(q-q^{-1})L'_0\,.
\eeq

Type 2 superalgebra consists of \eqref{CZtwist}, \eqref{GGtab2} and \eqref{LnGr2}:
\beq
[L'_n,L'_m]_{(m-n)}=[n-m]L'_{n+m}+a_{n,m}J_{n+m}\,, \qquad
[L'_n,J_m]_{(m-n)}=-q^{-n}[m]J_{n+m}\,,
\eeq
\beq
\{G_r,G_s\}_{(\frac{s-r}{2})}=q^{r+s+2}(q^{\frac{s-r}{2}}+q^{\frac{r-s}{2}})L'_{r+s}
+b_{r,s}J_{r+s}\,,
\eeq
\beq
[L'_n,G_r]_{(r-\frac{n}{2})}=q^{-n}[\frac{n}{2}-r]G_{n+r}\,, 
\eeq
where $a_{n,m}$ given by \eqref{a_nm}, and note that $a'=-b'=\frac{-1}{\qqi}$ (see \eqref{CZTgen}). 

%%%%%%%%%%%
%  Type 3   
%%%%%%%%%%%
\subsection{Type 3 super CZ algebra}

Type 3 was studied in \cite{superCZ}. 
The general form of Virasoro counterpart $L'_n$ is defined by \eqref{CZTgen} with 
the notation \eqref{Bn}, and 
we choose the coupling $g'_n$ as \eqref{Gtype3} for Type 3 super CZ algebra:
\beq
 L'_n=B_n-g'_n J_n\,,\qquad g'_n=q^{-\frac{n+1}{2}}[\frac{n+1}{2}] \,.
\eeq
The anomaly situation is the same as Type 2, and the bosonic part algebra is denoted as $\,\scz$. 
The supercharge $G_r$ is defined by \eqref{Gtype3}:
\beq
G_r=\lambda^{-\frac{1}{2}}x^{r+\frac{1}{2}}(\partial_\theta-\theta\partial_x)\,. 
\eeq
Choosing  $a=\frac{-1}{2}$ in \eqref{lam}, 
$\lambda$ is related to the $\,\scz$ zero mode $L'_0$ by \eqref{lambda}:
\beq
\lambda=\mu+(q^{-1}-1)\theta\delt=1+(\qqi)L'_0\,.
\eeq

Type 3 superalgebra consists of \eqref{CZtwist}, \eqref{GrGs2} and \eqref{LnGr2}:
\beq
[L'_n,L'_m]_{(m-n)}=[n-m]L'_{n+m}+a_{n,m}J_{n+m}\,, \qquad
[L'_n,J_m]_{(m-n)}=-q^{-n}[m]J_{n+m}\,,
\eeq
\beq
\{G_r,G_s\}_{(\frac{s-r}{2})}=q^{r+s+\frac{5}{2}}(q^{\frac{s-r}{2}}+q^{\frac{r-s}{2}})L'_{r+s}\,,
\eeq
\beq
[L'_n,G_r]_{(r-\frac{n}{2})}=q^{-n}[\frac{n}{2}-r]G_{n+r}\,, 
\eeq
where $a_{n,m}$ given by \eqref{a_nm}, 
and note that $a'=-q^{-1}b'=\frac{-q^{-1}}{\qqi}$ (see \eqref{CZTgen}). 

%%%%%%%%%%%%%%%%%%%%%%%%%%%%%%%%%%%%%%%%%%%%%%%%%%%%%%%%%%%%%%%%%%%
%                               Appendix  C
%%%%%%%%%%%%%%%%%%%%%%%%%%%%%%%%%%%%%%%%%%%%%%%%%%%%%%%%%%%%%%%%%%%
\setcounter{equation}{0}

\section{MT and MSB Representation}\label{app:MSB}

In this appendix, we summarize the realizations of the (super) CZ generators within the 
framework of MT operators, where $z$ denotes the one-dimensional parameter 
and $\partial$ its derivative operator. In the supersymmetric case, MT operators 
are embedded in MSB space.

Non-supersymmetric CZ algebras can be expressed in terms of two 
MT bases \eqref{BTz} and \eqref{JTz}:
\beq
\hat{B}_n=-\Op{T}{n}{0}=\frac{-z^n}{\qqi}\,,    \qquad
\hat{J}_n=q^{n+2\Delta}\Op{T}{n}{2}=z^n\frac{q^{-2z\partial}}{\qqi}\,. 
\eeq
The general forms of CZ and $\,\scz$ generators are given by
\begin{align}
& \hat{L}_n=\hat{B}_n-g_n \hat{J}_n\,,\qquad g_n=aq^{-2n}+b \,, \label{defcz1} \\
& \hat{L}'_n=\hat{B}_n-g'_n \hat{J}_n\,,\qquad g'_n=a'q^{-n}+b' \label{defcz2}\,.
\end{align}
The scaling operator $\hat{\mu}$ is given by \eqref{muhat}:
\beq
\hat{\mu}=1-(\qqi)z\partial_q=q^{-2z\partial} =c \hat{J}_0\,, 
\quad\mbox{where }\, c=\qqi\,.  
\eeq

The supersymmetric CZ generators (Type 3) are summarized in Section~\ref{sec:*form}. 
They are realized in MSB space, which is composed of the MT and spin matrix bases. 
The MSB representation of $CZ$ generators \eqref{supLmat} 
and deformed $U(1)$ current $\mathcal{\hat{J}}_n$ (denoted as $\mathcal{\hat{J}}_n^{'}$ in \eqref{newJ} with $w=1$) are:
\beq
\mathcal{\hat{L}}_n=q^{-2n\nu}\mathcal{\hat{L}}_n^+ \otimes\sigma_2\sigma_1
+\mathcal{\hat{L}}_n^- \otimes\sigma_1\sigma_2\,,
\eeq
\beq
\mathcal{\hat{J}}_n=qc\hat{J}_{n}\otimes\sigma_1\sigma_2\,,
\eeq
where component operators $\mathcal{\hat{L}}_n^\pm$ are given by \eqref{supLpm}:
\beq
\mathcal{\hat{L}}_n^+=\hat{B}_n+q^{-2n}\hat{J}_n\,,\qquad
\mathcal{\hat{L}}_n^-=\hat{B}_n+q^{-1-n}\hat{J}_n\,.
\eeq
It is straightforward to identify $\mathcal{\hat{L}}_n^+$ with $CZ$, and $\mathcal{\hat{L}}_n^-$ with $\,\scz$ generators, according to \eqref{defcz1} and \eqref{defcz2}. 

The supercharges are given by \eqref{Gmat}:
\beq
\mathcal{\hat{G}}_r=\mathcal{\hat{G}}_r^+ +\mathcal{\hat{G}}_r^-\,,\qquad
\mathcal{\hat{G}}_r^+=\hat{\mu}^{-\nu}\tilde{B}_r\otimes\sigma_1\,,\qquad
\mathcal{\hat{G}}_r^-=c\hat{\mu}^{\nu}\tilde{J}_r\otimes\sigma_2\,,
%\label{Gmat}
\eeq
where $\tilde{B}_r$ and $\tilde{J}_r$ are given by \eqref{BJtilde} 
with \eqref{abc_def} and \eqref{e1e2_def}:
\beq
\tilde{B}_r= q^{(1-2\nu)r+\nu+\frac{1}{2}} \hat{B}_{r-\frac{1}{2}}\,,\qquad
\tilde{J}_r=-q^{-r-\frac{1}{2}} \hat{J}_{r+\frac{1}{2}} - q^{r+\frac{1}{2}} \hat{B}_{r+\frac{1}{2}}\,.
\eeq

The superalgebras are given by \eqref{superLL}-\eqref{sumLL} for $N=1$, 
and \eqref{GGcom}-\eqref{hatJG} for $N=2$.

\medskip
\noindent
\textit{Remark.} The constant $c=q-q^{-1}$ appearing in the definition of $\hat{\mu}$ 
is simply absorbed into the notation of $\hat{J}_n$ and related operators in the 
subsequent formulas for convenience. Its elimination has no physical significance.

%%%%%%%%%%%%%%%%%%%%%%%%%%%%%%%%%%%%%%%%%%%%%%%%%%%%%%%%%%%%%%%%%%%
%                               Appendix  D
%%%%%%%%%%%%%%%%%%%%%%%%%%%%%%%%%%%%%%%%%%%%%%%%%%%%%%%%%%%%%%%%%%%
\setcounter{equation}{0}

\section{Weyl Matrix Representation}\label{app:Weyl}

In the Weyl matrix representation, the matrices $\mathrm{H}$ and 
$\mathrm{Q}$ (defined in \eqref{HQdef}) are used instead of MT operators. 
The counterparts of basic operators 
$\hat{B}_n$ and $\hat{J}_n$ are given by \eqref{BJmat}:
\beq
{\mathrm B}_n^\pm=\frac{\mp1}{\qqi}{\mathrm H}^n\,,\qquad
{\mathrm J}_n^\pm=\frac{1}{\qqi}{\mathrm H}^n{\mathrm Q}^{\pm2}\,. \label{eq6.8}
\eeq
The $CZ^\pm$ generators are realized as \eqref{LisBJ}:
\beq
{\mathrm L}^\pm_n={\mathrm B}_n^\pm + g_n^{\pm} {\mathrm J}_n^\pm\,,\qquad
g_n^{\pm}=\pm q^{\mp2n}\frac{1-A_n^{\pm}c}{\qqi}\,,
\eeq
where $A_n^{\pm}$ is defined in \eqref{QHCZ}.
$\mathrm{Q}^{\pm2}$ playing the role of scaling operators $\mathrm{S}_0^\pm$ 
are related to $\mathrm{L}_0^\pm$ as in \eqref{matS00}:
\beq
{\mathrm S}_0^\pm=1\pm(\qqi){\mathrm L}_0^\pm={\mathrm Q}^{\pm2}\,.
\eeq

The supersymmetric CZ generators (Type 3) are realized in tensor product space 
composed of Weyl and spin matrix bases. Explicit expressions for them 
in the case of $\nu=1$ are presented in Section~\ref{sec:TBM}; 
see \eqref{bbG+}-\eqref{bbJ}. 
Here we show the case of $\nu=\frac{1}{2}$ as another example. 

In the case of $\nu=\frac{1}{2}$, $\nu$-dependent parameters and operators (derived from \eqref{BJtilde}, \eqref{abc_def}, \eqref{p1def}, \eqref{matS00} and \eqref{mudiff}) are given by:
\beq
\gamma=1\,,\quad p_1=\frac{r+s}{2}\,,\quad
\hat{\mu}^{\pm\nu}\ra{\mathrm Q}^{\pm1}\,,\quad 
\tilde{B}_r\ra \tilde{{\mathrm B}}_r=
\frac{-q}{\qqi}{\mathrm H}^{r-\frac{1}{2}}\,.
\eeq
The remaining parts can be obtained through direct substitution $(\hat{B}_n,\hat{J}_n)\ra({\mathrm B}_n,{\mathrm J}_n)$ and application of the matrix representation \eqref{eq6.8}. 
By this substitution with \eqref{e1e2_def}, we redefine 
$(\mathcal{\hat{L}}_n,\mathcal{\hat{G}}_r^\pm,\mathcal{\hat{J}}_n)$ in terms of 
$(\mathbb{L}_n,\mathbb{G}_r^\pm,\mathbb{J}_n)$, 
yielding the matrix representation for each generator: 
\begin{align}
&\mathbb{G}_r^+=\frac{-q^{r+\frac{1}{2}}}{\qqi}{\mathrm H}^{r-\frac{1}{2}}
{\mathrm Q}^{-1}\otimes \sigma_1\,, \\
&\mathbb{G}_r^-=- {\mathrm H}^{r+\frac{1}{2}}
{\mathrm Q}(q^{-2r-1}{\mathrm Q}^2-1 ) \otimes \sigma_2\,,
\end{align}
\beq
\mathbb{L}_n=q^{-n}\mathbb{L}_n^+ \otimes\sigma_2\sigma_1
+\mathbb{L}_n^- \otimes\sigma_1\sigma_2\,, 
\eeq
where $\mathbb{L}_n^\pm$ and $\mathbb{J}_n$ are the same as those 
defined in \eqref{bbLpm} and \eqref{bbJ}. 
The superalgebras are given by \eqref{superLL}-\eqref{sumLL} for $N=1$, 
and \eqref{GGcom}-\eqref{hatJG} for $N=2$. 

%%%%%%%%%%%%%%%%%%%%%%%%%%%%%%%%%%%%%%%%%%%%%%%%%%%%%%%%%%%%%%%%%%%
%                               Appendix  E
%%%%%%%%%%%%%%%%%%%%%%%%%%%%%%%%%%%%%%%%%%%%%%%%%%%%%%%%%%%%%%%%%%%
\setcounter{equation}{0}

%%%%%%%%%%%%%%%%%%%%%%%%%%%%%%%%%%%%%
%  E    Electron system w/ Zeeman effect 
%%%%%%%%%%%%%%%%%%%%%%%%%%%%%%%%%%%%%
\section{Electron Spin System in Static Magnetic Field}\label{sec:SV}
\indent

The construction of superalgebras fundamentally requires Grassmann bases. To this end, we examine a one-electron spin system in a static magnetic field with spin-magnetic interaction (Zeeman term), which provides a natural quantum mechanical system with inherent Grassmann bases.
\beq
H=\frac{1}{2m}(\bm{\sigma}\cdot\bm{\pi})^2=\frac{1}{2m}\bm{\pi}^2
+\frac{1}{2}g \mu_B \bm{\sigma}\cdot \bm{B}\,.
\eeq
Here, $\sigma_i$, $\mu_B$, $g$ are the Pauli matrices, Bohr magneton, and $g$-factor respectively, where 
\beq
\mu_B=\frac{e\hbar}{2mc}\,,\quad g=2(1+\frac{\alpha}{2\pi}+O(\alpha^2))\,,\quad
\alpha=\frac{e^2}{\hbar c}
\eeq
($\alpha$ is the fine structure constant $\approx 1/137$). 
Neglecting relativistic effects, we set $g=2$ since this is not essential for the following discussion. Taking $\bm{B}=(0,0,B)$, we write
\beq
H=H_0 + \delta H\,,\quad H_0=\frac{1}{2m}\bm{\pi}^2 \,,\quad 
\delta H=\mu_B B\sigma_z \,.
\eeq
When there exists a base operator $\mathcal{O}_\beta$ (for example cyclotron center $\bm{\beta}$ or magnetic translation $\Tm{R}$) that commutes with $H_0$, the following construction forms bases that commute with $H$ (i.e., commute with $\sigma_z$):
\beq
\mathcal{O}_\beta\otimes 1\,,\quad \mathcal{O}_\beta\otimes\sigma_z\,,\quad
\mathcal{O}_\beta\otimes\sigma_1\sigma_2\,,\quad \mathcal{O}_\beta\otimes\sigma_2\sigma_1\,, \label{MSbase}
\eeq
where
\beq
\sigma_1\sigma_2=\begin{pmatrix}
0 & 0  \\
0 & 1  \\
\end{pmatrix} \,,\quad\quad
\sigma_2\sigma_1=\begin{pmatrix}
1 & 0  \\
0 & 0  \\
\end{pmatrix} 
\eeq
and $\sigma_1,\sigma_2$ constitute Grassmann bases that anticommute with $\delta H$ (i.e., anticommute with $\sigma_z$):
\beq
\{\sigma_1,\sigma_2\}=1\,,\quad \sigma_1^2=\sigma_2^2=0\,,
\eeq
where
\beq
\sigma_1=\sigma_x-i\sigma_y=\begin{pmatrix}
0 & 0  \\
1 & 0  \\
\end{pmatrix} \,,\quad\quad
\sigma_2=\sigma_x+i\sigma_y=\begin{pmatrix}
0 & 1  \\
0 & 0  \\
\end{pmatrix}\,.
\eeq
These Grassmann bases can be mapped to Grassmann variables and their derivatives in superspace $(x,\theta)$. We refer to this as the SSM correspondence (superspace and spin matrix):
\beq
\sigma_1\leftrightarrow \theta\,,\quad 
\sigma_2\leftrightarrow \partial_\theta\,.  \label{grass}
\eeq
While refs.~\cite{JS,KS} used this basis with magnetic translation (MT) to realize supersymmetric algebra of $q$-Virasoro algebra different from CZ algebra, super CZ algebra has been constructed primarily through quantum superspace (QSS)~\cite{superCZ,superCZ2}. Although its physical realization using MT has recently been achieved~\cite{SCZ4}, this paper provides a detailed theoretical foundation for understanding the correspondence between QSS-based super CZ algebra of ref.~\cite{superCZ} and its MT realization.

First, we review ordinary Virasoro superalgebra on ordinary superspace (SS) (subsection~\ref{sec:vsa}) and derive Virasoro superalgebra on SM space by applying the SSM correspondence \eqref{grass}. Similarly, we examine super CZ on QSS in Section~\ref{sec:SCZ} and investigate the extension of SSM correspondence to QSS version in Section~\ref{sec:MSB}.

%%%%%%%%%%%%%%%%%%%%%%%%%
%  E.1     super Virasoro
%%%%%%%%%%%%%%%%%%%%%%%%%
\subsection{Virasoro super algebra}\label{sec:vsa}
\indent

Here, we deal with a situation where the magnetic field is very weak, allowing us to take $q=1$, and the Virasoro superalgebra is realized in a state where only supersymmetry remains. First, we prepare the Virasoro operator $V_n$ and $U(1)$ operator $F_n$ as bosonic fundamental operators:
\beq
V_n=-x^{n+1}\partial_x\,,\quad  F_n=x^n\,.
\eeq
These satisfy the following Virasoro and $U(1)$ commutation relations:
\begin{align}
&[V_n,V_m]=(n-m)V_{n+m}\,,\quad [V_n,F_m]=-m F_{n+m}\,, \label{BBBF} \\
& [F_n,F_m]=0\,.  \label{FFU1}
\end{align}
The composite operator
\beq
 L^B_n=V_n-\frac{n+1}{2}F_n\,
\eeq
and $F_n$ also satisfy the same algebras as \eqref{BBBF} and \eqref{FFU1},
\begin{align}
&[L^B_n,L^B_m]=(n-m)L^B_{n+m}\,,\quad[L^B_n,F_m]=-mF_{n+m}\,,\quad \\
&[L^B_n,V_m]=(n-m)L^B_{n+m}-\frac{m(m+1)}{2}F_{n+m}\,.
\end{align}

Let $J_n$ denote the super current obtained by applying the operator $F_n$ (which represents a component of the dilatation on bosonic space) to the superspace $(x,\theta)$. Then, involving the super Virasoro operator $L_n$ and the supercharge $G_r$, we have
\begin{align}
&L_n=V_n-\frac{n+1}{2}J_n\,,  \label{EYform1} \\
&J_n=F_n\theta\partial_\theta\,,\quad G_r=x^{r+\frac{1}{2}}(\partial_\theta-\theta\partial_x)\,,
\end{align}
and the Virasoro super algebra ($N=1$) is realized:
\begin{align}
&[L_n,L_m]=(n-m)L_{n+m}\,,\quad \{G_r,G_s\}=2L_{r+s}\,, \label{SVfrom} \\
&[L_n,G_r]=(\frac{n}{2}-r)G_{n+r}\,, \\
&[L_n,J_m]=-mJ_{n+m}\,,\quad [J_n,J_m]=0\,.
\end{align}
The decomposition into $N=2$ super Virasoro algebra is given by
\beq
G_r=G_r^+ + G_r^-\,,\quad G_r^-=x^{r+\frac{1}{2}}\partial_\theta\,,\quad 
G_r^+=-x^{r+\frac{1}{2}} \theta \partial_x\,, 
\eeq
satisfying the following relations:
\begin{align}
&\{G_r^\pm,G_s^\pm\}=0\,,\quad \{G_r^+,G_s^-\}=L_{r+s}+\frac{1}{2}(r-s)J_{r+s}\,, \\
&[L_n,G_r^\pm]=(\frac{n}{2}-r)G_{n+r}^\pm\,, \quad[J_n,G_r^\pm]=\pm G_{n+r}^\pm\,. \label{SVto}
\end{align}

By applying the SSM correspondence \eqref{grass} to $G_r,J_n,L_n$, we can obtain their MSB representation
\begin{align}
&\mathcal{G}_r=V_{r-\frac{1}{2}}\otimes\sigma_1 + F_{r+\frac{1}{2}}\otimes\sigma_2\,,
\label{Gform2} \\
&\mathcal{J}_n=F_n\otimes\sigma_1\sigma_2\,, \label{Jform2} \\
&\mathcal{L}_n=V_n\otimes1 -\frac{n+1}{2}F_n\otimes\sigma_1\sigma_2\,,
\label{EYform2}
\end{align}
and their explicit matrix representations are as follows:
\beq
\mathcal{G}_r=\mathcal{G}_r^+ + \mathcal{G}_r^-\,,\quad
\mathcal{G}_r^+=\begin{pmatrix}
0 &  0  \\
V_{r-\frac{1}{2}} & 0  \\
\end{pmatrix} \,,\quad\quad
\mathcal{G}_r^-=\begin{pmatrix}
0 &  F_{r+\frac{1}{2}}  \\
0 & 0  \\
\end{pmatrix} \,,
\eeq
\beq
\mathcal{J}_n=\begin{pmatrix}
0 & 0  \\
0 & F_n  \\
\end{pmatrix} \,,\quad\quad
\mathcal{L}_n=\begin{pmatrix}
V_n & 0  \\
0 & L^B_n  \\
\end{pmatrix} \,,\quad\quad \label{BL2let}
\eeq
These $\mathcal{L}_n,\mathcal{G}_r,\mathcal{G}_r^\pm, \mathcal{J}_n$ satisfy the above Virasoro super algebra ($N=1,2$) \eqref{SVfrom}-\eqref{SVto}.

Readers should note that, while some notation is deliberately kept the same to indicate the correspondence between the Virasoro and CZ systems, the content of this appendix is independent of the main text and should not be confused with the main discussion.

%%%%%%%%%%%%%%%%%%%%%%%%%%%%%%%%%%%%%%%%%%%%%%%%%%%%%%%%%%%%%%%%%%%

%\newpage
\newcommand{\NP}[1]{{\it Nucl.{}~Phys.} {\bf #1}}
\newcommand{\PL}[1]{{\it Phys.{}~Lett.} {\bf #1}}
\newcommand{\Prep}[1]{{\it Phys.{}~Rep.} {\bf #1}}
\newcommand{\PR}[1]{{\it Phys.{}~Rev.} {\bf #1}}
\newcommand{\PRL}[1]{{\it Phys.{}~Rev.{}~Lett.} {\bf #1}}
\newcommand{\PTP}[1]{{\it Prog.{}~Theor.{}~Phys.} {\bf #1}}
\newcommand{\PTPS}[1]{{\it Prog.{}~Theor.{}~Phys.{}~Suppl.} {\bf #1}}
\newcommand{\MPL}[1]{{\it Mod.{}~Phys.{}~Lett.} {\bf #1}}
\newcommand{\IJMP}[1]{{\it Int.{}~J.{}~Mod.{}~Phys.} {\bf #1}}
\newcommand{\IJTP}[1]{{\it Int.{}~J.{}~Theor.{}~Phys.} {\bf #1}}
\newcommand{\JPA}[1]{{\it J.{}~Phys.} {\bf A}:\ Math.~Gen. {\bf #1}~}
\newcommand{\JHEP}[1]{{\it J.{}~High Energy{}~Phys.} {\bf #1}}
\newcommand{\JMP}[1]{{\it J.{}~Math.{}~Phys.} {\bf #1} }
\newcommand{\CMP}[1]{{\it Commun.{}~Math.{}~Phys.} {\bf #1} }
\newcommand{\LMP}[1]{{\it Lett.{}~Math.{}~Phys.} {\bf #1} }
\newcommand{\doi}[2]{\,\href{#1}{#2}\,}  %Note: _ to be escaped by {\_} in #2

%%%%%%%%%%%%%%%%%%%%%%%%%%%%%%%%%%%%%%%%%%%%%%%%%%%%%%%%%%%%%%%%%%%%%%%
%                           REFERENCES       
%%%%%%%%%%%%%%%%%%%%%%%%%%%%%%%%%%%%%%%%%%%%%%%%%%%%%%%%%%%%%%%%%%%%%%%

\end{document}